\documentclass{aa}
\usepackage{natbib}
\usepackage[dvips,final]{graphics}
\usepackage{psfrag}

\psfrag{l}{$\lambda$}
\psfrag{W}{$\Omega$}
\psfrag{p}{$p$}
\psfrag{F}{$F$}
\psfrag{H}{$($}
\psfrag{D}{$D$}
\psfrag{\310}{$|$}
\psfrag{L}{$)$}
\psfrag{=}{$=$}
\psfrag{,}{$,$}
\psfrag{0.3}{$0.3$}

\begin{document}

\thesaurus{02(12.07.1; 12.03.4; 12.03.3)}

\title{
Gravitational lensing statistics with extragalactic surveys
}

\subtitle{
I.  A lower limit on the cosmological constant
}

\author{
Ralf Quast\inst{1} 
\and 
Phillip Helbig\inst{2}
} 

\institute{
Universit{\"a}t Hamburg, 
Hamburger Sternwarte, 
Gojenbergsweg 112, 
D-21029 Hamburg, 
Germany
\and 
University of Manchester, 
Nuffield Radio Astronomy Laboratories, 
Jodrell Bank, 
Macclesfield,
Cheshire SK11 9DL, 
England}

\date{Received 2 December 1998 / Accepted 14 January 1999}

\offprints{R.~Quast}
\mail{rquast@hs.uni-hamburg.de}

\authorrunning{R.~Quast \& P.~Helbig}
\titlerunning{
Gravitational lensing statistics with extragalactic surveys. I
}

\maketitle

\begin{abstract}
We reanalyse optical gravitational lens surveys from the literature in
order to determine relative probabilities in the
$\lambda_{0}$-$\Omega_{0}$ plane, using a softened singular isothermal
sphere lens model.  In addition, we examine a portion of the
$\lambda_{0}$-$\Omega_{0}$ plane which includes all viable cosmological
models; this is vital for comparison with other cosmological tests.  The
results are, within the errors, consistent with those of more
specialised analyses, such as those concerning upper limits on
$\lambda_{0}$ in a flat universe.  We note that gravitational lensing
statistics can provide a quite robust \emph{lower} limit on the
cosmological constant as well, which could prove important in confirming
current claims of a positive cosmological constant.  At 95\% confidence,
our lower and upper limits on $\lambda_{0}-\Omega_{0}$, using lens
statistics information alone, are respectively $-3.17$ and $0.3$.  For a
flat universe, these correspond to lower and upper limits on
$\lambda_{0}$ of respectively $-1.09$ and $0.65$. 

\end{abstract}

\section{Introduction}

The use of gravitational lensing statistics as a cosmological tool was
first considered in detail by \citet{ETurnerOG84a}; the influence of the
cosmological constant was investigated thoroughly by
\citet{MFukugitaFKT92a}, building on the work of \citet{ETurner90a} and
\citet{MFukugitaFK90a}.  More recently, \citet[hereafter K96, and
references therein]{CKochanek96a} and \citet{EFalcoKM98a} have laid the
groundwork for using gravitational lensing statistics for the detailed
analysis of extragalactic surveys.  However, these analyses either have
concentrated on a small subset of the possible cosmological models as
described by the density parameter $\Omega_{0}$ and the cosmological
constant $\lambda_{0}$, have used a simpler (singular) lens model or
both.  This analysis is the first time $\lambda_{0}$ and $\Omega_{0}$
have been used as independent parameters in conjunction with a
non-singular lens model in an analysis of this type, complementing
similar analyses with other emphases.  (See \citet{YChengMKrauss99a} for
a discussion of the importance of including a core radius.) Also, we
include enough of the $\lambda_{0}$-$\Omega_{0}$ plane to avoid
neglecting any possibly viable models; this also makes the comparison
with a variety of other cosmological tests easier.  This is especially
important in light of the fact that many analyses
\citep[e.g.][]{SPerlmutteretal98a,ARiessetal98a,BSchmidtetal98a,
RCarlbergetal98a,CLineweaver98a,EGuerraDW98a,RDalyGW98a} are now
suggesting that our universe may contain a significant cosmological
constant \emph{and} be non-flat. 

The plan of this paper is as follows.  Sect.~\ref{theory} reviews the
groundwork and serves to define our notation.  In Sect.~\ref{obsprior}
we specify the observational data and selection functions we use and
formulate prior information about the parameters $\lambda_0$ and
$\Omega_0$.  Sect.~\ref{calculations} describes the parametric submodels
we use and the numerical computations we perform.  In Sect.~\ref{results}
we discuss our results and compare them with others.
Sect.~\ref{conclusions} presents our summary and conclusions.

\section{Probability of multiply imaged sources}
\label{theory}

In this section we briefly review the statistical concepts introduced in
K96; this also serves to define our notation.  Note that with regard to
cosmogical notation we follow that of \citet{RKayserHS97a}, repeating
here only 2 equations needed for discussion in this paper: the comoving
spherical volume element at redshift $z$ reads 
\begin{equation}
  \mathrm{d}V =
  4\pi r^2\frac{c}{H_0} \frac{\mathrm{d}z}{\sqrt{Q(z)}},
  \label{eq:vol}
\end{equation}
where
\begin{equation}
\label{eq:Q}
  Q(z) = 
  \Omega_0(1 + z)^3 - (\Omega_0 + \lambda_0 - 1)(1 + z)^2 + \lambda_0.
\end{equation}

Following the K96 approach, we assume that the light deflection
properties of the gravitational lenses can be modelled with a particular
type of circularly symmetric lens models with a monotonically declining
radial mass profile.  Such lens models generally create three images and
have two critical radii on which the magnification diverges
\citep[e.g.][]{PSchneiderEF92a}.  It is possible to estimate the
probability $p(m,z_{\mathrm{s}})$ of the event 
\begin{quote}  
\em
  A source at redshift $z_{\mathrm{s}}$ is triply imaged.  The total 
  apparent magnitude of the three images is $m$.  The image 
  configuration meets the selection criteria $S$ and, particularly, 
  shows the properties $C$.
\end{quote} 
If the outer and inner critical angular radii of the lens potential are
respectively $r_+$ and $r_-$, the image magnification at radial angular
position $r$ is $\mu(r)$, the total magnification of the three images of
a source at angular position $y$ is $M(y)$, the functions $S(y)$ and
$C(y)$ are $0|1$ valued selection functions, the comoving density of
lenses of luminosity $L$ is $\mathrm{d}n/\mathrm{d}L$ and the
number-magnitude counts of sources are $\mathrm{d}N/\mathrm{d}m$, then 
\begin{eqnarray}
  p(m,z_{\mathrm{s}}) & = & 
  \frac{1}{2}
  \int\limits_0^{z_{\mathrm{s}}} \frac{\mathrm{d}V}{\mathrm{d}z}
    \int\limits_0^\infty \frac{\mathrm{d}n}{\mathrm{d}L}
      \int\limits_{r_-}^{r_+} r|\mu(r)|^{-1}
\qquad\times\nonumber\\
& & 
\times\qquad
        \underbrace{B(m,z_{\mathrm{s}},y)S(y)C(y)}
      \,\mathrm{d}r
    \,\mathrm{d}L
  \,\mathrm{d}z,
  \label{eq:p}
\end{eqnarray}
where
\begin{equation}
\label{eq:bias}  
  B(m,z,y) = 
  \frac{\mathrm{d}N}{\mathrm{d}m}\{m +2.5 \log[M(y)], z\}
  \left[
    \frac{\mathrm{d}N}{\mathrm{d}m}(m, z)
  \right]^{-1}.
\end{equation}
The critical radii, the image magnifications and the source position are
functions of the lens model, the luminosity of the lens galaxy and the
redshifts of the source and the lens galaxy.  If the underbraced
functions are dropped, Eq.~(\ref{eq:p}) yields the optical depth -- the
fraction of the sky included within the caustics of all lenses between
us and the sources at redshift $z_{\mathrm{s}}$.  The inclusion of these
functions accounts for magnification bias, survey selection effects
(including what is defined as a lensing event) and allows the observed
image separation to be taken into account. 

Equation~(\ref{eq:p}) parametrically depends on $\lambda_0$ and
$\Omega_0$ through Eq.~(\ref{eq:vol}) and through the angular size
distances,\footnote{In general, the angular size distances depend not
only on $\lambda_{0}$ and $\Omega_{0}$ but on the degree of homogeneity
in the universe as well \citep[see, e.g.,][]{RKayserHS97a}.  However, in
contrast to some other cosmological tests, this effect is relatively
unimportant for the type of analysis performed here \citep[see,
e.g.,][]{MFukugitaFKT92a}.} which are needed for calculating observable
quantities from the lens model (these also depend on the source and lens
redshifts).  Equation~(\ref{eq:p}) additionally depends on parametric
submodels required to model the lens population and the number-magnitude
counts of sources.  Since throughout this paper we are principally
interested in $\lambda_0$ and $\Omega_0$, hereafter we refer to the
submodel parameters as nuisance parameters (although technically they
are on the same footing with $\lambda_{0}$ and $\Omega_{0}$, there are
not of as much interest here and thus a nuisance).  In principle, one
could also incorporate other observables into the parametric model; the
reasons for not doing so are practical. 

Assuming the survey selection function $S$ is known, we can numerically
compute Eq.~(\ref{eq:p}) and reasonably estimate the probability
$1-p(m_i,z_i)$ that the quasar $i$ is singly imaged or the probability
$p(m_i,z_i,\theta_i)$ that the quasar $i$ is multiply imaged and its
images (within some tolerance) are separated by $\theta_i$.  If the
survey data $D$ contains $M$ singly and $N$ multiply imaged quasars, we
can estimate the probability of the event 
\begin{quote}  
  \em
  In a model universe fixed by the cosmological parameters $\lambda_0$,
  $\Omega_0$ and the nuisance parameters $\vec{\xi}$, a multiply imaged 
  quasar survey collects the observational data~$D$.  
\end{quote}
by applying the parametric model (or likelihood function)
\begin{eqnarray}
\label{eq:lf}
  \ln[p(D|\Omega_0,\lambda_0,\vec{\xi})] & = & 
  -\sum_{i=1}^{M} p(m_i,z_i) 
\nonumber\\
& & + 
  \sum_{j=1}^{N} \ln[p(m_j,z_j,\theta_j)],
\end{eqnarray}
where the logarithm $\ln[1-p(m_i,z_i)]$ was expanded to first order.  We
can combine surveys of different objects by adding the logarithms of the
likelihood functions for the individual surveys, and can combine surveys
containing the same objects by applying their joint selection function. 

In Bayesian theory the model parameters $\lambda_0$, $\Omega_0$,
$\vec{\xi}$ are regarded as random quantities with known joint prior
probability density function $p(\lambda_0,\Omega_0,\vec{\xi})$.  Applying
Bayes's theorem, the appropriate posterior probability distribution
given the observational data $D$ is 
\begin{equation}
  p(\lambda_0,\Omega_0\vec{\xi}|D) =
  p(D|\lambda_0,\Omega_0,\vec{\xi}) \otimes p(\lambda_0,\Omega_0,\vec{\xi}),
\end{equation}
where the operation `$\otimes$' denotes multiplication followed by
normalisation.  Marginalising the nuisance parameters 
\begin{equation}
\label{eq:marg}
  p(\lambda_0,\Omega_0|D) =
  \int\limits p(\lambda_0,\Omega_0,\vec{\xi}|D) \,\mathrm{d}\vec{\xi}.
\end{equation}
yields the (marginal) posterior probability density function for the
parameters $\lambda_0$ and $\Omega_0$.  In the limit where all nuisance
parameters take a precise value, $\vec{\xi}=\vec{\xi}_0$, the joint
prior probability density function $p(\lambda_0,\Omega_0,\vec{\xi})$
factorises into $p(\lambda_0,\Omega_0)$ and a delta distribution
$\delta(\vec{\xi}-\vec{\xi}_0)$, and Eq.~(\ref{eq:marg}) simplifies to 
\begin{equation}
\label{eq:simplemarg}
  p(\lambda_0,\Omega_0|D) =
  p(D|\lambda_0,\Omega_0,\vec{\xi}_0) \otimes p(\lambda_0,\Omega_0).
\end{equation}
On the basis of Eq.~(\ref{eq:marg}) or Eq.~(\ref{eq:simplemarg}), we can
calculate confidence regions for two parameters or perform further
marginalisations and calculate mean values, standard deviations and
marginal confidence intervals for one parameter.

\section{Observational data and prior information}
\label{obsprior}

We use the observational data of the optical multiply imaged quasar
surveys by \citet{DCramptonMF92b}, \citet{AJaunsenJPS95a},
\citet{CKochanekFS95a}, \citet{HYeeFT93a} and the observational data of
the HST Snapshot Survey compiled by \citet{DMaozBSBDDGKMY93a}, including
\mbox{\object{Q 0142-100}}, \mbox{\object{Q 1115+080}} and
\mbox{\object{Q~1413+117}}.  If applicable, we replace the apparent
quasar V magnitude catalog data found in \citet{DCramptonMF92b},
\citet{AJaunsenJPS95a} and \citet{HYeeFT93a} with more current data from
\citet{MVeronCettyPVeron96a}.  We estimate the \citet{CKochanekFS95a}
apparent quasar V magnitude data by adding the survey average V--R and
V--I colours to the observational R and I magnitude data.  Following
K96, we only include quasars with redshift~$z_{\mathrm{s}}>1$.  In all,
our sample contains~807 singly and~5 multiply imaged quasars.  The
observational data of the multiply imaged quasars are summarised in
Table~\ref{ta:data}.  Our complete input data can be obtained from 
\begin{quote}
\verb|http://multivac.jb.man.ac.uk:8000/ceres|\\
                                       \verb|/data_from_papers/lower_limit.html|
\end{quote}
\begin{table}
   \caption[]{
      Observational data of multiply imaged quasars contained in the
      sample.  The magnitudes are V magnitudes unless otherwise
      specified.  The image separations are taken from 
      \citet{CKochanekFILMR97a}}
   \label{ta:data}
   \begin{tabular*}{\linewidth}{@{\extracolsep{\fill}}llll}
      \hline
      \hline
      Identifier & $m$ [mag] & $z_{\mathrm{s}}$ &
            $\theta$ [$\arcsec$]  \\
      \hline
      \object{Q 0142$-$100}  & $17.0$   & $2.72$ & $2.2$  \\
      \object{Q 1009$-$0252} & $18.1$ B & $2.74$ & $1.5$  \\
      \object{Q 1115$+$080}  & $16.2$   & $1.72$ & $2.2$  \\
      \object{Q 1208$+$1011} & $17.9$   & $3.80$ & $0.48$  \\
      \object{Q 1413$+$117}  & $17.0$   & $2.55$ & $1.2$  \\
      \hline
   \end{tabular*}
\end{table}
This follows K96 for purposes of comparison.  Since much larger surveys
(i.e.~CLASS) will be considered in a future paper, there is little point
in increasing the number of lenses for its own sake.  Since radio
observations are considered in more detail in a companion paper
\citep{PHelbigMQWBK99a}, we restrict ourselves to optical surveys in
this paper.  We use the \citet{DCramptonMF92b}, HST Snapshot Survey and
\citet{HYeeFT93a} survey selection functions proposed in
\citet{CKochanek93c}, the \citet{AJaunsenJPS95a} survey selection
function at $1\farcs0$ seeing and the preliminary \citet{CKochanekFS95a}
survey selection function. 

Before considering prior information in more detail, one must first
decide which region of the $\lambda_{0}$-$\Omega_{0}$ plane is to be
investigated.  Clearly, this region should be defined by either exact
constraints or conservative estimates, as opposed to current `best fit'
values (and their errors), in order to avoid excluding any possibly
viable cosmological models.  Also, it is desirable for the region to be
on the large side, so that in addition the sensitivity of the test
(i.e.~what regions of the $\lambda_ {0}$-$\Omega_{0}$ plane can be ruled
out at a high confidence level) can be investigated.

\subsection{The range of $\Omega_{0}$ and $\lambda_{0}$}

The mass clustered with galaxies on smaller scales,
$\Omega_{0,\mathrm{gal}}$, is 0.1 within a factor of two
\citep[e.g.][]{PPeebles93a}.  This lower limit is small compared to our
full $\Omega_{0}$ range so we do not assume any prior lower limit on
$\Omega_{0}$ except, of course, 
\begin{equation}
\Omega_{0} \ge 0.
\end{equation}
Especially for comparison with other work it is important to note that,
within the framework of cosmological models based on general relativity
with which we (and almost everyone else at present) are working,
$\Omega_{0} \ge 0$ is a \emph{requirement}.  Results reported which
include $\Omega_{0} < 0$ within the errors, or even as a best-fit value,
do not indicate `implausible results' but merely improper statistics.
Often, confidence contours are assumed to be ellipses and these are
extended, if applicable, to $\Omega_{0} < 0$.  (Of course, it is
possible that $\Omega_{0} = 0$ is within the errors or even the best fit
value for a certain set of results.) 

An extremely conservative upper limit comes from dynamical tests on
larger (though still cosmologically small) scales; when this work was
started, we assumed an (again, extremely conservative) upper limit
$\Omega_0\le2$ \citep{OCzoske95a}. Since then, these methods have
started to indicate smaller values of $\Omega_{0}$,
\citep[e.g.][]{LdCostaNFGHSW97a} more in line with both a long tradition
of low $\Omega_{0}$ values
\citep[e.g.][]{JGottGST74a,PColesGEllis94a,PColesGEllis97a} (albeit with
somewhat larger errors) as well as new determinations (often with quite
small errors), examples of which are mentioned in Sect.~\ref{olprior}. 

We have assumed no prior upper or lower limits on $\lambda_{0}$ per se.
This has two reasons: 
\begin{itemize} 
\item `Direct' measurements of $\lambda_{0}$ (as opposed to 
measurements of a combination of parameters
involving $\lambda_{0}$) are virtually nonexistent. 
\item We obtain a small enough range in $\lambda_{0}$ from the values 
obtained from joint constraints on the range of $\Omega_{0}$ and 
$\lambda_{0}$. 
\end{itemize} 

Historically, positive $\lambda_{0}$ values have been considered more
than negative ones, probably because positive values can have a wide
range of relatively easily observable effects, while negative ones are
more difficult to measure.  Many cosmological tests have a degeneracy
such that $\lambda_{0}$ and $\Omega_{0}$ are correlated, so that
increasing $\lambda_{0}$ can be compensated for in some sense by
increasing $\Omega_{0}$ as well.  Thus, effects of negative values of
$\lambda_{0}$ for a given value of $\Omega_{0}$ are hard to
differentiate from the effects of larger values of $\Omega_{0}$ for
larger (less negative) values of $\lambda_{0}$ or even $\lambda_{0}=0$. 

Here, we consider negative values of $\lambda_{0}$ as well.  There is no
a priori reason why they cannot exist.  \emph{If} one believes that the
`source' of $\lambda_{0}$ are zero-point fluctuations of a quantum
vacuum, this would lend support to the idea that $\lambda_{0}>0$.
However, it is not clear that this \emph{must} be the \emph{only} source
of $\lambda_{0}$, and indeed it has been argued that, if this source of
$\lambda_{0}$ exists, there must be an additional contribution with a
\emph{negative} value \citep[e.g.][though the assumption that this is
possible is so obvious to the authors it is barely
stated!]{HMartelSW98a}. 

In spatially closed ($k=+1$) models, the antipode is required to be at
$z > 4.5 $, the redshift of the most redshifted multiply imaged object
currently known \citep{JGottPL89a,MParkJGott97a}.\footnote{Recently, a
lensed object of even larger redshift has been detected at $z=4.92$
\citep{MFranxvDIKT98a}.  However, at our resolution this would make only
a negligible difference to the results so we have not updated the
calculations to reflect this.} The light grey shaded area in the panel
in the middle of the left column of Fig.~\ref{fi:prior} marks the right
side of the region thus enclosed.  This gives us a slightly
$\Omega_{0}$-dependent upper limit on $\lambda_{0}$ which is slightly
stronger than that obtained by merely excluding models with no big bang.
(This can be done because these models have a maximum redshift which is
less than the redshift of high-redshift objects, the only exception
being some cosmological models which have $\Omega_{0} < 0.05$, the
robust lower limit discussed above \citep[e.g.][]{BFeige92a}.) 

The age of the universe in units of the Hubble time, $H_{0}^{-1}$, is
\begin{equation}
  \tau_0 = \int\limits_0^\infty \frac{\mathrm{d}z}{(1+z)\sqrt{Q(z)}}
  \label{eq:tau},
\end{equation}
where $Q(z)$ is given by Eq.~(\ref{eq:Q}) and thus depends on
$\Omega_{0}$ and $\lambda_{0}$.  (There are world models in which the
maximum redshift is not infinite but these are all models without a big
bang and are excluded by the constraint from the antipodal redshift or
the lower limit on $\Omega_{0}$ as discussed above and are thus not
relevant for this work.)  Clearly, in any physically realistic world
model, $\tau_0H_{0}^{-1}$ exceeds the age of the oldest galactic
globular clusters: 
\begin{equation}
   \tau_0 > t_{\mathrm{gc}}H_0.
  \label{eq:ineq}
\end{equation}
Following \citet{SCarrollPT92a}, we take a robust lower limit on
$\lambda_{0}$ from conservative lower limits on the Hubble constant and
age of the universe.  This gives a lower limit on $\lambda_{0}$ from the
value at $\Omega_{0} = 0$; at larger values of $\Omega_{0}$ the
constraint on $\lambda_{0}$ is not as strict---by assuming the lower
limit of $\lambda_{0} = -5$ independent of $\Omega_{0}$ we are being
conservative.  We choose $\lambda_{0}\ge -5$ instead of $\lambda_{0}\ge
-7$ as in \citet{SCarrollPT92a} since no published current constraints
examine this region in detail.  (Were this the case, then including this
area would be helpful if only to aid a direct comparison.)  This value
corresponds roughly to the \emph{one-sided} 99\% confidence level in the
top row of Fig.~\ref{fi:prior} (see Sect.~\ref{olprior}), which is also
a reason not to extend the area to more negative $\lambda_{0}$ values.

\subsection{Prior probability for $\lambda_{0}$ and $\Omega_{0}$}
\label{olprior}

We have assumed no prior knowledge of $\lambda_{0}$ per se, apart from
the upper and lower limits discussed above.  This has three reasons: 
\begin{itemize}
\item `Direct' measurements of $\lambda_{0}$ (as opposed to measurements
of a combination of parameters involving $\lambda_{0}$) are virtually
nonexistent. 
\item Based on general knowledge from the literature and our own
low-resolution calculations, we expect lens statistics itself to
constrain $\lambda_{0}$ quite well. 
\item Although recent measurements are encouraging (see
Sect.~\ref{results}), the value of $\lambda_{0}$ is observationally not
as well established as that of $\Omega_{0}$.
\end{itemize}

Regarding $t_{\mathrm{gc}}$ and $H_0$ as independent random quantities
with known prior probability density functions $p(t_{\mathrm{gc}})$ and
$p(H_0)$, the probability that Eq.~(\ref{eq:ineq}) is satisfied is 
\begin{equation}
  P(\tau_0 > t_{\mathrm{gc}}H_0) =
  \int\limits_0^\infty p(H_0) \int\limits_0^{\tau_0/H_0} p(t_{\mathrm{gc}})
  \,\mathrm{d}t_{\mathrm{gc}}\,\mathrm{d}H_0.
  \label{eq:P}
\end{equation}
A cosmological world model is compatible with the absolute age of the
oldest galactic globular clusters as long as the above expression does
not vanish.  Reasonably, we assume a prior probability density function
that is proportional to this expression 
\begin{equation}
  p_1(\lambda_0,\Omega_0) = 1 \otimes
  \int\limits_0^\infty p(H_0) \int\limits_0^{\tau_0/H_0} p(t_{\mathrm{gc}})
  \,\mathrm{d}t_{\mathrm{gc}}\,\mathrm{d}H_0.
  \label{eq:p1}
\end{equation}
The best estimate of the absolute age of the oldest galactic globular
clusters currently is $t_{\mathrm{gc}}=11.5\pm1.3\,\mbox{Gyr}$
\citep{BChaboyerDKK98a}.  We choose to formulate this prior information
in the form of a lognormal distribution that meets these statistics 
\begin{equation}
  p(t_{\mathrm{gc}}) =
  L(t_{\mathrm{gc}}|11.5\,\mbox{Gyr}, 1.3\,\mbox{Gyr}).
  \label{eq:ptgc}
\end{equation}
Similarly, we roughly estimate
$H_0=65\pm10\,\mbox{km}\,\mbox{s}^{-1}\,\mbox{Mpc}^{-1}$ and choose to
formulate this prior information in form of a normal distribution 
\begin{equation}
  p(H_0) =
  N(H_0|65\,\mbox{km}\,\mbox{s}^{-1}\,\mbox{Mpc}^{-1},
    10\,\mbox{km}\,\mbox{s}^{-1}\,\mbox{Mpc}^{-1}),
  \label{eq:ph0}
\end{equation}
where the notation for $L$ and $N$ is such that the two arguments
correspond to the mean and standard deviation. 

This estimate is compatible with `small' values of the Hubble constant,
which is conservative in the sense that it restricts our region of the
$\lambda_{0}$-$\Omega_{0}$ plane less than would `large' values.  By the
same token we neglect any time between the big bang and the formation of
the oldest globular clusters.  Inserting Eq.~(\ref{eq:ptgc}) and
Eq.~(\ref{eq:ph0}) in Eq.~(\ref{eq:p1}) one obtains a well-founded a
priori probability distribution for the parameters $\Omega_0$ und
$\lambda_0$. 

Although observational evidence has always indicated a low value of
$\Omega_{0}$
\citep[e.g.][]{JGottGST74a,PColesGEllis94a,PColesGEllis97a}, the
inflationary paradigm \citep[e.g.][]{AGuth81a}, coupled with a prejudice
against a non-negligible value of $\lambda_{0}$, has created a prejudice
in favour of $\Omega_{0} = 1$,\footnote{After this was found to conflict
with too many observations, the prejudice against a non-negligible value
of $\lambda_{0}$ weakened, and the new prejudice has been in favour of a
flat universe with $\lambda_{0} + \Omega_{0} = 1$.} unfortunately too
often to the extent where this prior belief has been elevated to the
status of dogma \citep[see, e.g.,][for an illuminating
account]{DMatraversES95a} even though there are serious fundamental
problems with the inflationary idea \citep[e.g.][]{RPenrose89a} and even
though there might be other solutions to the problems it claims to solve
\citep[e.g.][]{JBarrow95a,GCollins97a}. What is more, some current
inflationary thinking \citep[e.g.][]{NTurokSHawking98a} seems able to
predict values for $\lambda_{0}$ and $\Omega_{0}$ similar to current
observationally determined values, though it would have been more
interesting had this prediction been made before the recent improvements
in the observational situation.  (To be fair, many leading practitioners
of inflation consider a flat universe to be a robust prediction and its
observational falsification essentially a falsification of the entire
paradigm.)  Recently, in the light of overwhelming observational
evidence in favour of a low value of $\Omega_{0}$
\citep[e.g.][]{RCarlbergetal97a,RCarlberg98a,RCarlbergetal98b,
NBahcall97a,NBahcallFC97a,XFanBC97a,MBartelmannHCJP98a, CLineweaver98a},
whether determined more or less independently or in combination with
other parameters, this prejudice is starting to weaken.  Conservatively,
these results can be summarised as 
\begin{equation}
  p_2(\lambda_0,\Omega_0) =
  L(\Omega_0|0.4,0.2).
  \label{eq:p2}
\end{equation}
A prior constraint on $\Omega_{0}$ is useful since lensing statistics
alone, as expected and as our results show, cannot usefully constrain
$\Omega_{0}$. 

In addition, we also consider the product of $p_2(\lambda_0,\Omega_0)$
with the age constraint $p_1(\lambda_0,\Omega_0)$, 
\begin{equation}
  p_3(\lambda_0,\Omega_0) =
  p_1(\lambda_0,\Omega_0) \otimes p_2(\lambda_0,\Omega_0).
  \label{eq:p3}
\end{equation}

\subsection{General discussion of prior information}

\begin{figure*}
   \hfill
   \resizebox{0.375\textwidth}{!}{\includegraphics{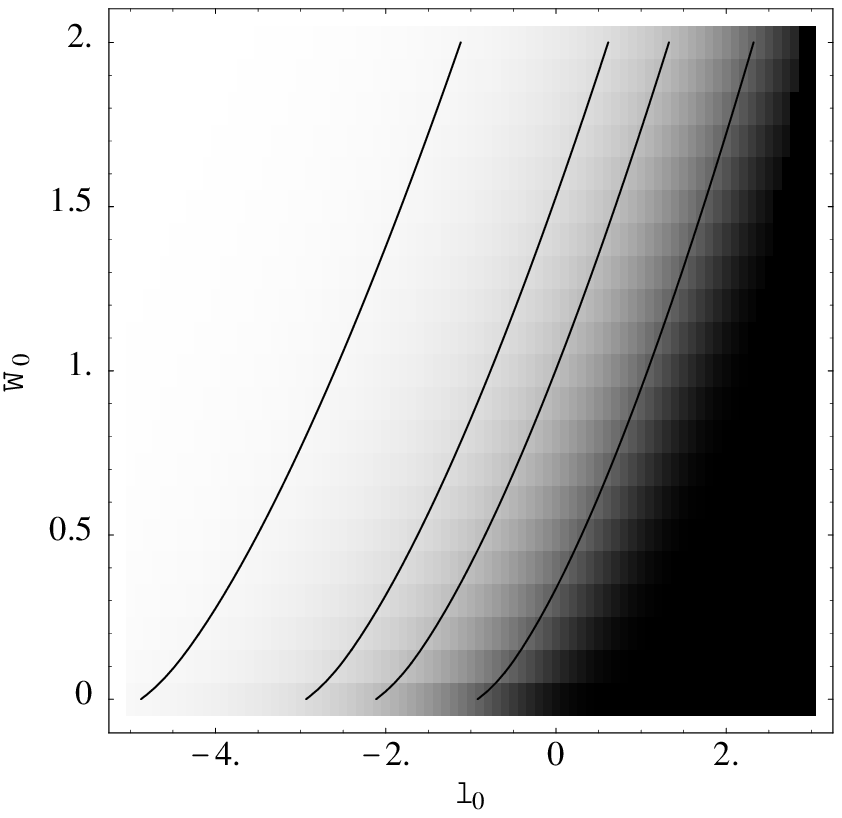}}

   \noindent
   \resizebox{0.375\textwidth}{!}{\includegraphics{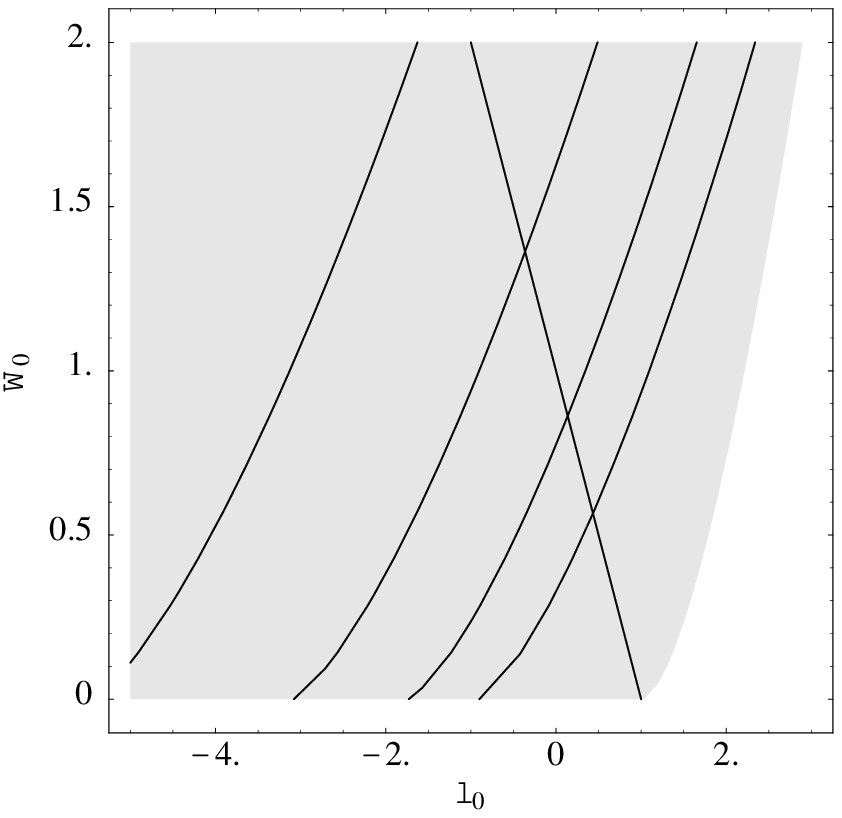}}
   \hfill
   \resizebox{0.375\textwidth}{!}{\includegraphics{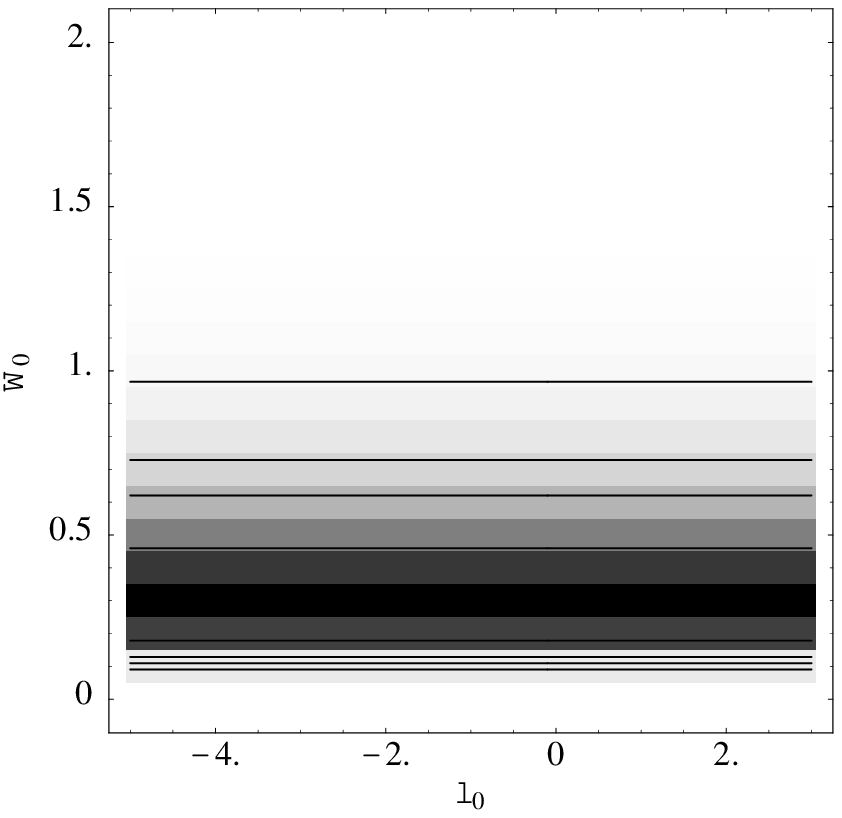}}

   \noindent
   \hfill
   \resizebox{0.375\textwidth}{!}{\includegraphics{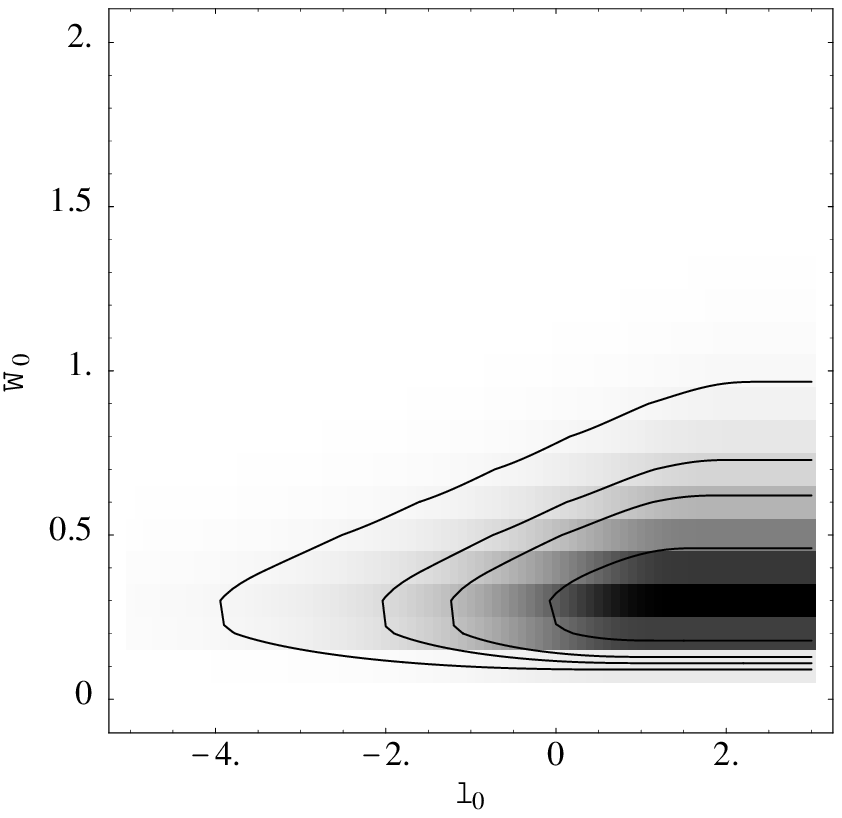}}
   \caption[]{
      \emph{Left column:}
      The cosmological parameter plane.  The four curved lines in the
      plot at the left are the isochrones \mbox{$t_0H_0=0.5,\dots,0.8$}.
      The straight line marks spatially flat world models.  In the
      white region, the antipodal redshift falls below $z=4.5$, the
      redshift of the most redshifted multiply imaged object currently
      known \citep{JGottPL89a,MParkJGott97a}. 
      \emph{Right column:}
      The prior probability distributions $p_1(\lambda_0,\Omega_0)$ (top
      panel), $p_2(\lambda_0,\Omega_0)$ (middle panel) and
      $p_3(\lambda_0,\Omega_0)$ (bottom panel).  The pixel grey level is
      directly proportional to the probability density ratio, darker
      pixels reflect higher ratios.  The pixel size reflects the
      resolution of our numerical computations.  The contours mark 0.61,
      0.26, 0.14 and 0.036 of the peak likelihood for the parameters
      $\lambda_0$ and~$\Omega_0$, which would correspond to the
      boundaries of the minimum $0.68$, $0.90$, $0.95$ and $0.99$
      confidence regions \emph{if the distribution were Gaussian}} 
   \label{fi:prior}
\end{figure*}

Using harsher constraints would mean that results would reflect almost
exclusively the prior information as opposed to the information derived
from lensing statistics.  It is not the purpose of this paper to do a
joint analysis of several cosmological tests,\footnote{but see
Sect.~\ref{results}} but rather to examine lens statistics as a
cosmological test.  For practical reasons, an upper limit on
$\Omega_{0}$ and upper and lower limits on $\lambda_{0}$ are required.
On the other hand, it is sensible to combine the results with
conservative constraints from other well-understood cosmological tests
where there is general agreement and little room for debate.  Within our
upper and lower limits, we present our results both with and without the
constraints discussed above.  The density values and confidence contours
of the three prior probability density functions are shown in the right
column of Fig.~\ref{fi:prior}. 

\section{Calculations}
\label{calculations}

Following K96, we use the \citet{GHinshawLKrauss87a} softened isothermal
sphere model for modeling the light deflection properties of the lens
galaxies.  For this model, the lens equation reads 
\begin{equation}
\label{eq:lens}
  x - y =
  \frac{bx}{\hat{s} + \sqrt{x^2 + \hat{s}^2}},
\end{equation} 
where $x$ is the angular position in the lens plane, $y$ the 
angular position in the 
source plane, $b\equiv4\pi(\sigma/c)^2(D_{\mathrm{ds}}/D_{\mathrm{os}})$,
$\sigma$ denotes the one--dimensional velocity dispersion of the dark
matter, $s$ denotes the core radius, $\hat{s}\equiv s/D_{\mathrm{od}}$ 
is the angular core radius 
and $D_{\mathrm{od}}$, $D_{\mathrm{os}}$ and $D_{\mathrm{ds}}$ denote
the angular size distances between the observer and the lens galaxy, the
observer and the source and the lens galaxy and the source,
respectively.  Still following K96, we model the distribution of
elliptical and lenticular lens galaxies using Schechter functions with
constant comoving density
\begin{equation}
  n_{\mathrm{e}}=0.61\pm0.21\,h^3\,10^{-2}\,\mbox{Mpc}^{-3}
\end{equation}
($h=H_0\,10^{-2}\,\mbox{km}^{-1}\,\mbox{s}\,\mbox{Mpc}$) and slope
\begin{equation}
\alpha_{\mathrm{e}}=-1.0\pm0.15.
\end{equation}
The lens galaxy luminosities are
converted to the dark matter velocity dispersions of the softened
isothermal lens model by means of Faber--Jackson type relations,
\begin{equation}
L/L_{*\mathrm{e}}=(\sigma/\sigma_{*\mathrm{e}})^{\gamma_{\mathrm{e}}},
\end{equation}
where 
\begin{equation}
\gamma_{\mathrm{e}}=4.0\pm0.5
\end{equation}
and
\begin{equation}
\sigma_{*\mathrm{e}}=225.0\pm22.5\,\mathrm{km\,s}^{-1}.
\end{equation}
The core radii of the softened isothermal lens model are varied with the
dark matter velocity dispersions according to \begin{equation}
s/s_{*\mathrm{e}}=(\sigma/\sigma_{*\mathrm{e}})^{2+\varepsilon},
\end{equation} where $\varepsilon=2.8$ and
$s_{*\mathrm{e}}=10h^{-1}\,\mbox{pc}$.  We consider elliptical and
lenticular lens galaxies only.  For the number--magnitude counts of
quasars, we adopt the best-fit model from K96.  We neglect here
evolution, dust and other possible systematic effects and refer the
reader to K96 for a discussion. 

In our first calculations we apply Eq.~(\ref{eq:simplemarg}) and compute
the a priori likelihood
\begin{equation}
p(D|\lambda_0,\Omega_0,\vec{\xi}_0) 
\end{equation}
and the posterior probability density functions 
\begin{equation}
  p_1(\lambda_0,\Omega_0|D) =
  p(D|\lambda_0,\Omega_0,\vec{\xi}_0) \otimes p_1(\lambda_0,\Omega_0),
  \label{eq:simplemarg1}
\end{equation}
\begin{equation}
  p_2(\lambda_0,\Omega_0|D) =
  p(D|\lambda_0,\Omega_0,\vec{\xi}_0) \otimes p_2(\lambda_0,\Omega_0),
  \label{eq:simplemarg2}
\end{equation}
and
\begin{equation}
  p_3(\lambda_0,\Omega_0|D) =
  p(D|\lambda_0,\Omega_0,\vec{\xi}_0) \otimes p_3(\lambda_0,\Omega_0)
  \label{eq:simplemarg3}
\end{equation}
in the limit where all nuisance parameters take precisely their mean
values.  To obtain an impression of the consequences of neglecting the
uncerntainties of the nuisance parameters, in our second calculation we
increase the value of the most uncertain nuisance parameter,
$n_{\mathrm{e}}$, by two standard deviations. 

For the computation of the innermost integral on the right side of
Eq.~(\ref{eq:p}), we consider the detectability of images in pairs: If
the separation between the two closest images -- these are always images
2 and 3, counting from the outside in -- is more than the lower limit of
the survey resolution limit $S(y)$, we define the image separation and
flux ratio for the purpose of sample selection based on the two
brightest images, usually 1 and 2.  Otherwise we construct one image
from the combined fluxes and flux-weighted positions of images 2 and 3 and
define the image separation and flux ratio for the purpose of sample
selection based on this combination image and image~1. 

In general, if the separation between images 1 and 2 is too large for
the survey \emph{and} the separation between images 2 and 3 is large
enough, then the image separation and flux ratio for the purpose of
sample selection should be based on images 2 and 3. However, the present
surveys are sensitive to the largest separations due to isolated
galaxies, so this case doesn't need to be addressed in this paper
(i.e.~implementing it would lead to the same results in the present
case). 

For the calculation of the probabilities $p(m_i,z_i,\theta_i)$ the
function $C(y)$ selects only those image configurations whose separation
is $\pm10$ per cent of the observed separation $\theta_i$.

Each of the three integrals on the right side of Eq.~(\ref{eq:p}) is
approximated to an accuracy better than $0.004$ by a family of recursive
monotone stable formulae \citep{PFavatiLR91a,PFavatiLR91b}.

\section{Results and discussion}
\label{results}

\subsection{Information content}

Given some observational data $D$, some model parameters $\vec{\phi}$,
and some prior and posterior probability density functions
$p(\vec{\phi})$ and $p(\vec{\phi}|D)$, the amount of information
obtained from the data \citep[e.g.][]{JBernardoASmith94a}
(on a logarithmic scale) is 
\begin{equation}
  \log[I(D)] =
  \int\limits p(\vec{\phi}) \log
  \left[
    \frac{p(\vec{\phi}|D)}{p(\vec{\phi})}
  \right]
  \,\mathrm{d}\vec{\phi}.
  \label{eq:info}
\end{equation}
The amounts of information obtained from our sample data are given in
the caption of Fig.~\ref{fi:posterior}.

\subsection{Results}

\begin{figure*}
   \resizebox{0.375\textwidth}{!}{\includegraphics{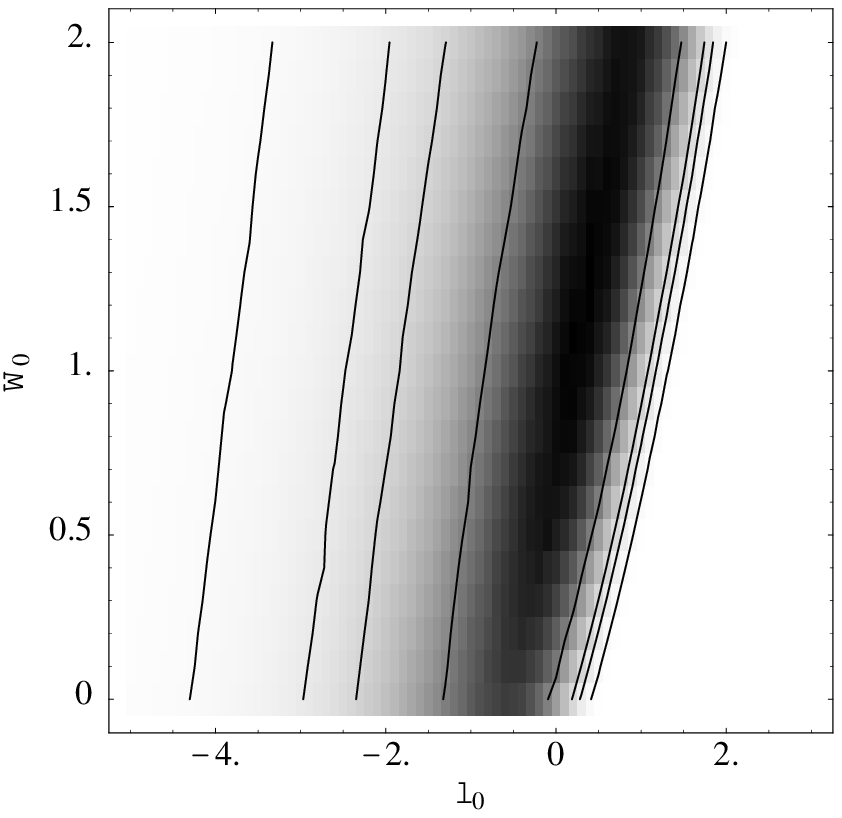}}
   \hfill
   \resizebox{0.375\textwidth}{!}{\includegraphics{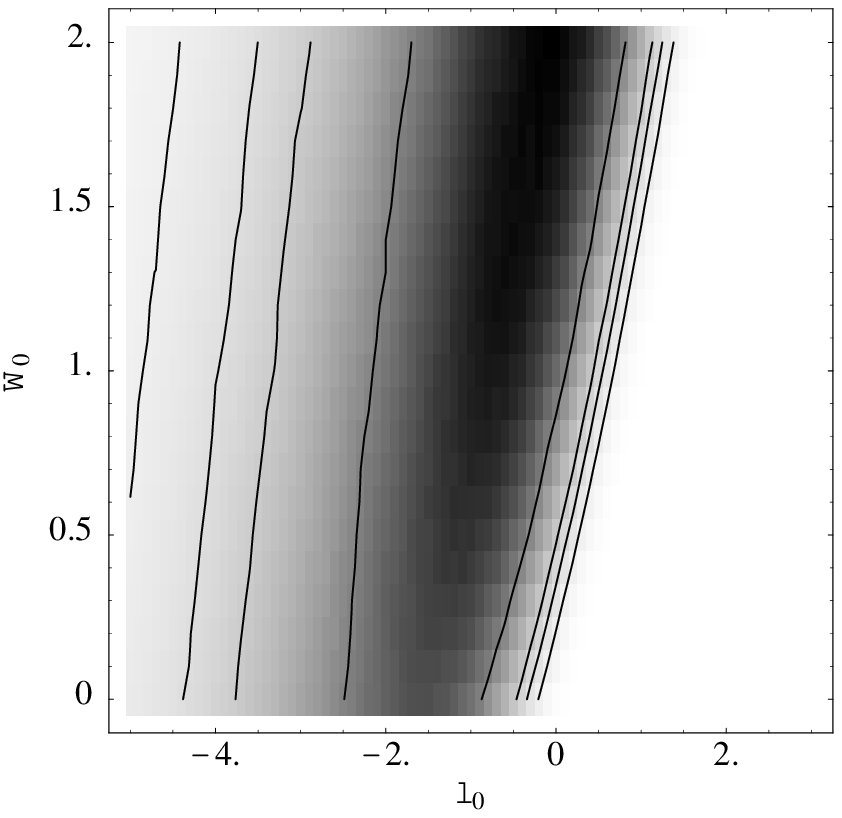}}
   \caption[]{
      \emph{Left panel:} The likelihood function
      $p(D|\lambda_0,\Omega_0,\vec{\xi}_0)$.  All nuisance parameters
      are assumed to take precisely their mean values.  The pixel grey
      level is directly proportional to the likelihood ratio, darker
      pixels reflect higher ratios.  The pixel size reflects the
      resolution of our numerical computations.  The contours mark the
      boundaries of the minimum $0.68$, $0.90$, $0.95$ and $0.99$
      confidence regions for the parameters $\lambda_0$ and $\Omega_0$. 
      \emph{Right panel:}
      Exactly the same as the left panel, but the parameter 
      $n_{\mathrm{e}}$ is increased by two standard deviations}
   \label{fi:likelihood}
\end{figure*}
The left panel of Fig.~\ref{fi:likelihood} shows the constraints on the
cosmological parameters $\lambda_{0}$ and $\Omega_{0}$ based only on the
information obtained from the lens statistics. 

Quite good constraints can be placed on $\lambda_{0}$, more or less
independent of $\Omega_{0}$.
It is a
well-known fact (see K96 and references therein) that lensing statistics
can provide a good \emph{upper} limit on $\lambda_{0}$.  While in the
past this has mainly been discussed in the context of flat cosmological
models, it is of course more general \citep{SCarrollPT92a,EFalcoKM98a}.
Although no unexpected effects are seen, it is important to note that
this is the first time $\lambda_{0}$ and $\Omega_{0}$ have been used as
independent parameters in conjunction with a non-singular lens model in
an analysis of this type. 

Our analysis shows for the first time that gravitational lensing
statistics can place a quite firm \emph{lower} limit on $\lambda_{0}$ as
well, again more or less independent of $\Omega_{0}$.  The constraint is
not as tight since the gradient in the probability density is not as
steep towards negative $\lambda_{0}$ as towards positive $\lambda_{0}$.
If this lower limit can be improved enough, it could provide an
independent confirmation of the detection of a positive cosmological
constant (see Sect.~\ref{compare}).  On the other hand, this might be
difficult, since Poisson errors in the number of lenses and
uncertainties in the normalisation of the luminosity density of galaxies
introduce relatively large uncertainties in this region of parameter
space \citep[K96,][]{EFalcoKM98a}.  The latter effect is illustrated in
the right panel of Fig.~\ref{fi:likelihood}, where $n_{\mathrm{e}}$, the galaxy
luminosity density normalisation, is increased by two standard deviations: 
the derived
lower limit on $\lambda_{0}$ changes much more than does the upper limit.
Nevertheless, our robust lower limit is much better than the $-7$
mentioned in \citet{SCarrollPT92a}. 

Our results place no useful constraints on $\Omega_{0}$.  It is
interesting to note the fact, however, that likely values of
$\lambda_{0}$ and $\Omega_{0}$ are positively correlated.  This is
similar to most cosmological tests, a notable exception being
constraints derived from CMB anisotropies (see Sect.~\ref{compare}). 
Fortunately, constraints on $\Omega_{0}$ from other sources are quite 
good (Sect.~\ref{olprior}).  Often, this is cast in the form of a 
constraint on $\Omega_{0} - \lambda_{0}$ \citep[e.g.][]{ACoorayQC99a}
or, perhaps more practical, $\lambda_{0} - \Omega_{0}$.  This is a 
reasonable way or reducing the information to one number, at least 
when one is concerned with upper limits on $\lambda_{0}$ (or 
$\lambda_{0} - \Omega_{0}$) in a relatively low-density universe.  
Besides the obvious dependencies on confidence levels and assumptions 
made, when comparing constraints on $\lambda_{0}$ from different 
investigations one should keep in mind whether they are approximations,
like $\lambda_{0} - \Omega_{0}$ in lensing statistics, and whether a 
value for a particular scenario (for example, for a flat universe) is 
the `obvious' definition or in fact describes the intersection of the 
$k=0$ line with the corresponding 2-dimensional confidence contour, 
which in general will give a different number.  Also, some authors plot 
`real' confidence contours while some actually plot contours at values 
which would correspond to certain confidence contours were the 
likelihood distribution in the parameter space in question Gaussian.

\begin{figure*}
   \resizebox{0.375\textwidth}{!}{\includegraphics{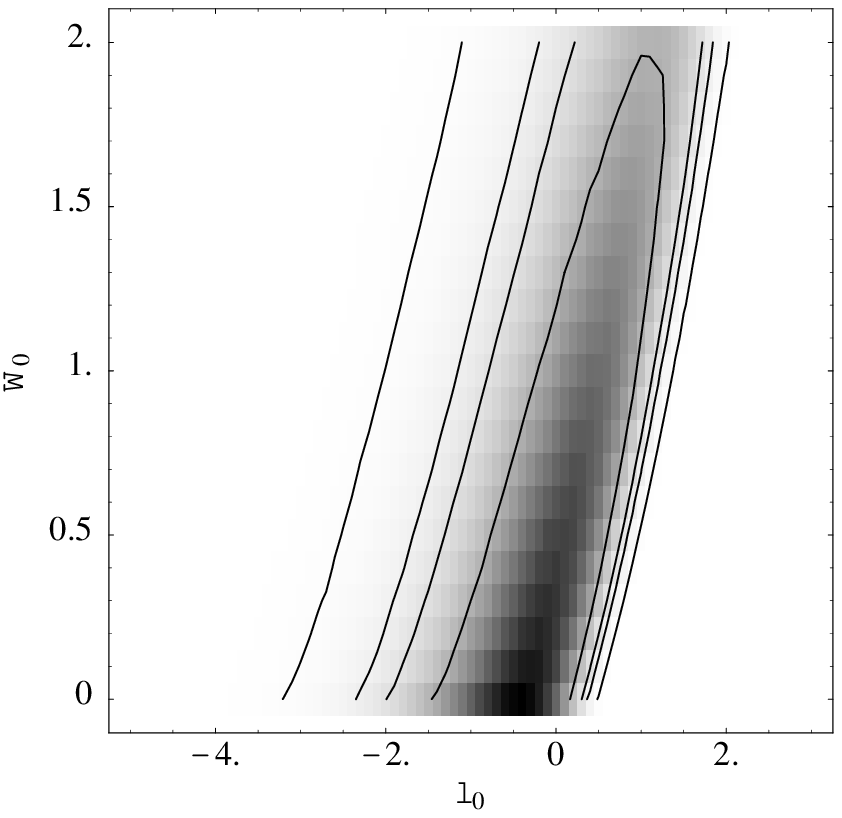}}
   \hfill
   \resizebox{0.375\textwidth}{!}{\includegraphics{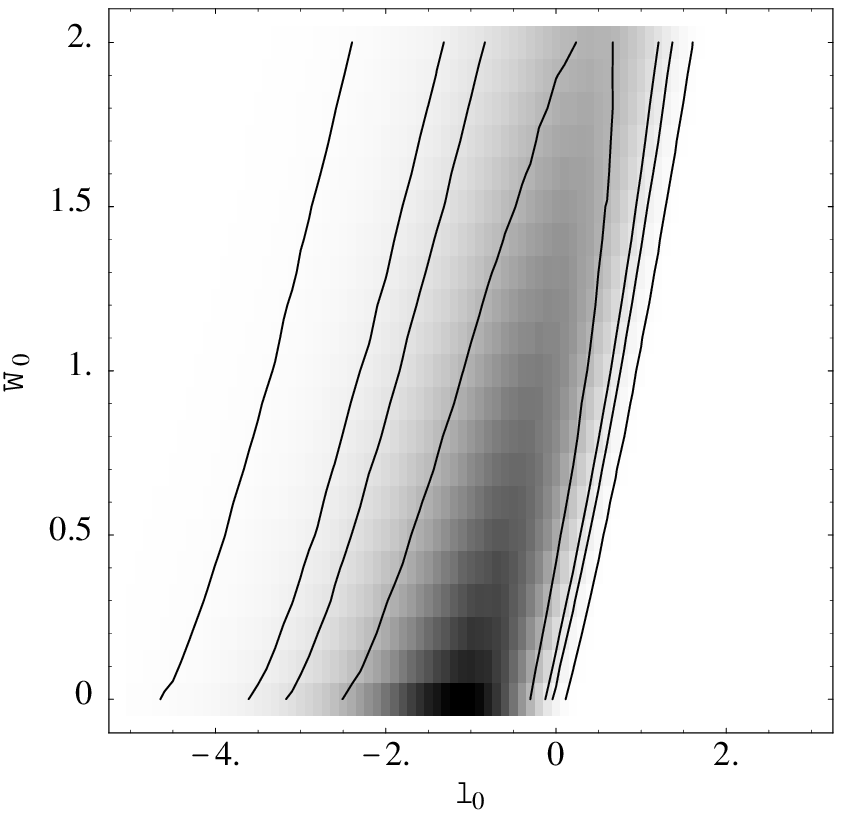}}

   \noindent
   \resizebox{0.375\textwidth}{!}{\includegraphics{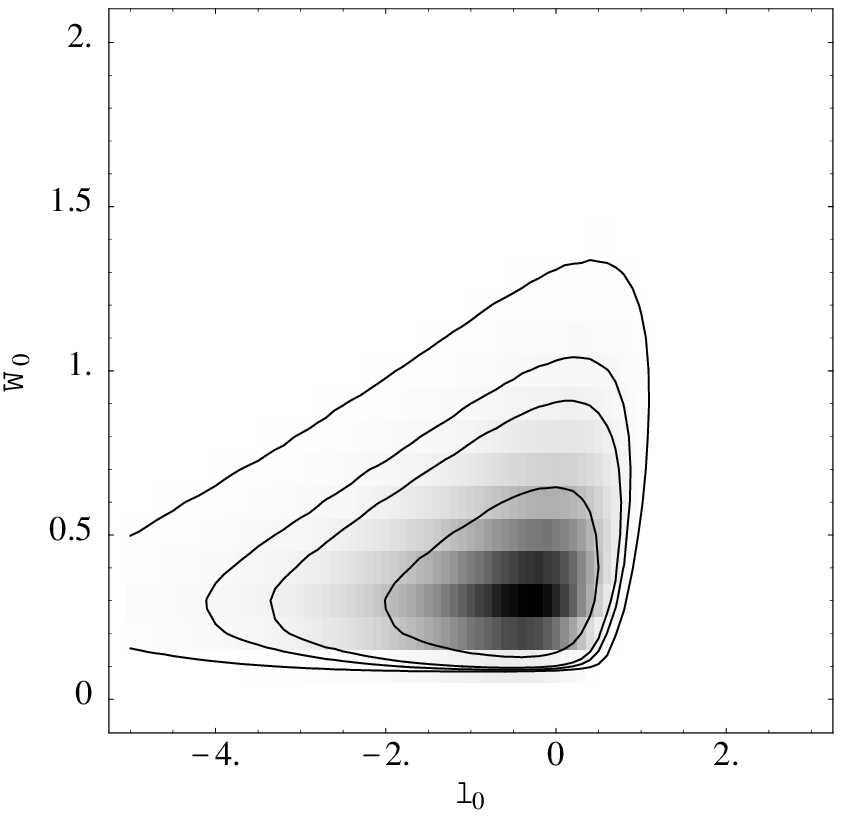}}
   \hfill
   \resizebox{0.375\textwidth}{!}{\includegraphics{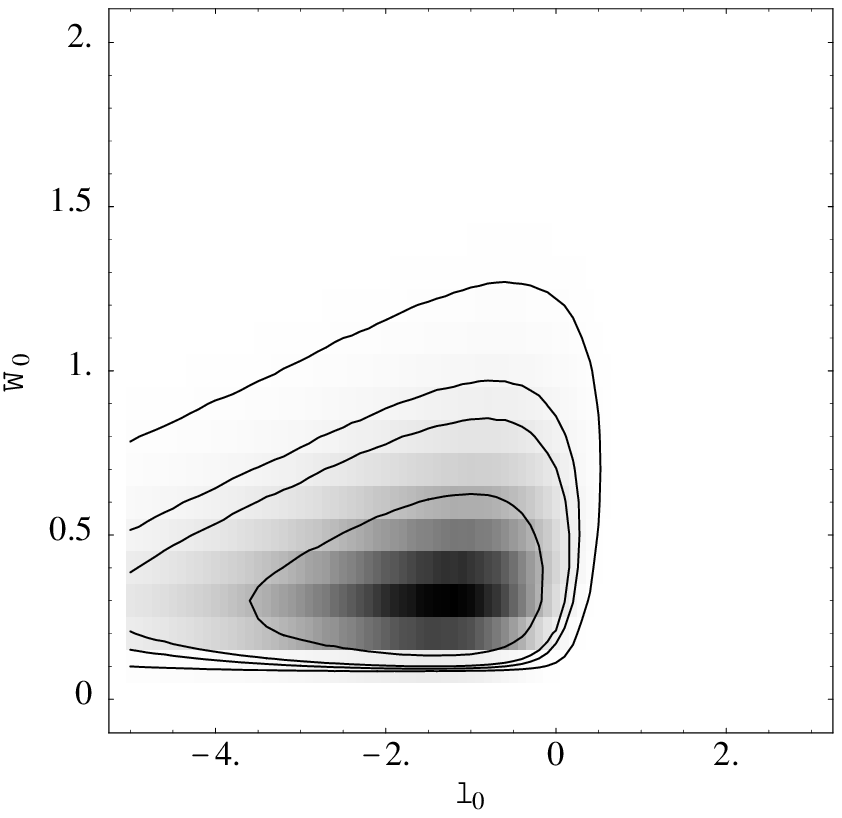}}

   \noindent
   \resizebox{0.375\textwidth}{!}{\includegraphics{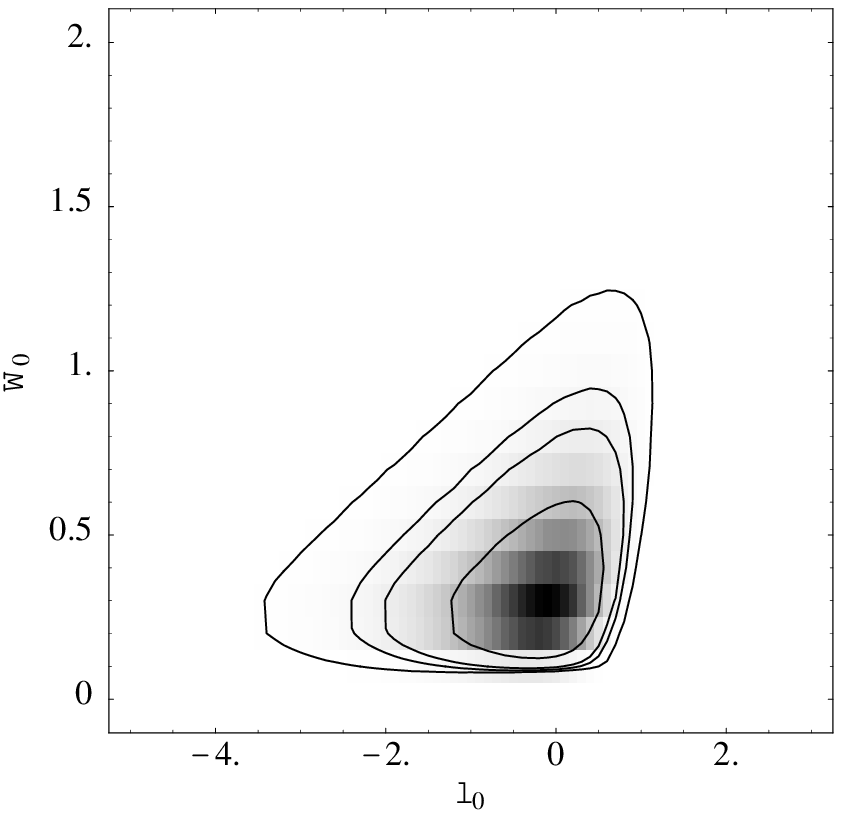}}
   \hfill
   \resizebox{0.375\textwidth}{!}{\includegraphics{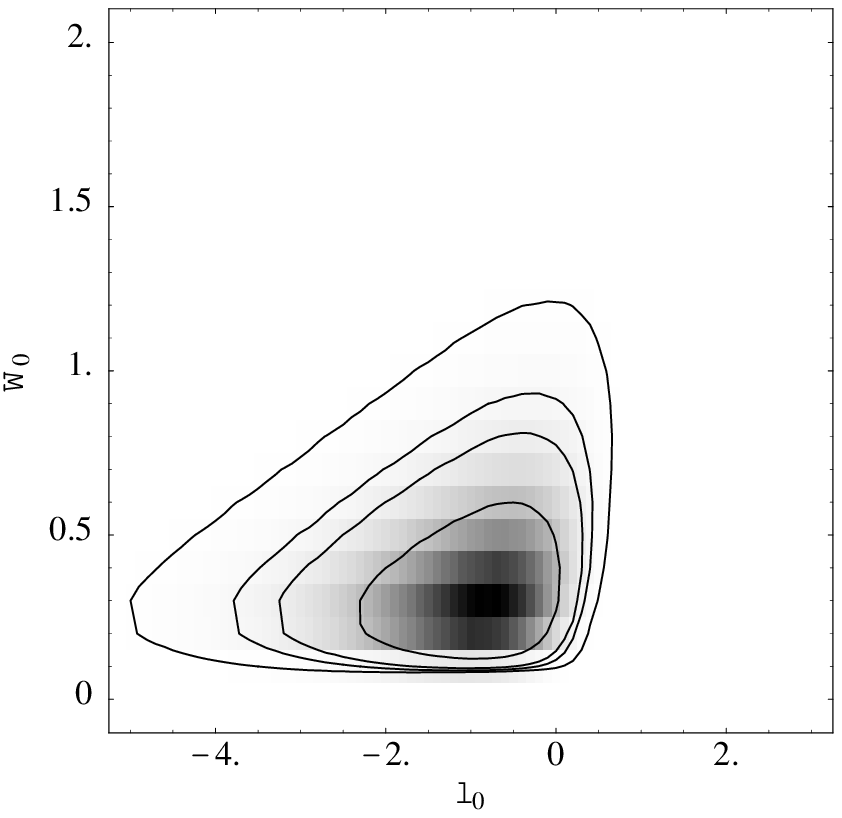}}
   \caption[]{
      \emph{Left column:} The posterior probability density functions
      $p_1(\lambda_0,\Omega_0|D)$ (top panel),
      $p_2(\lambda_0,\Omega_0|D)$ (middle panel) and
      $p_3(\lambda_0,\Omega_0|D)$ (bottom panel).  All nuisance
      parameters are assumed to take precisely their mean values.  The
      pixel grey level is directly proportional to the likelihood ratio,
      darker pixels reflect higher ratios.  The pixel size reflects the
      resolution of our numerical computations.  The contours mark the
      boundaries of the minimum $0.68$, $0.90$, $0.95$ and $0.99$
      confidence regions for the parameters $\lambda_0$ and $\Omega_0$.
      The respective amounts of information (Eq.~(\ref{eq:info}))
      obtained from our sample data are $I_1=1.74$, $I_2=1.24$ and
      $I_3=1.74$.  
      \emph{Right column:} Exactly the same as the left column, but the 
      parameter $n_{\mathrm{e}}$ is increased by two standard deviations} 
   \label{fi:posterior}
\end{figure*}
The left plot in the top row of Fig.~\ref{fi:posterior} shows the
joint likelihood of our lensing statistics analysis and that
obtained by using conservative estimates for $H_{0}$ and the age
of the universe (see Sect.~\ref{olprior}).  Although neither
method alone sets useful constraints on $\Omega_{0}$, their
combination does, since the constraint involving $H_{0}$ and the
age of the universe only allows large values of $\Omega_{0}$ for
$\lambda_{0}$ values which are excluded by lens statistics.  Even
though the 68\%~contour still allows almost the entire
$\Omega_{0}$ range, it is obvious from the grey scale that much
lower values of $\Omega_{0}$ are favoured by the joint
constraints.  The upper limit on $\lambda_{0}$ changes only
slightly while, as is to be expected, the lower limit becomes
tighter.  Also, the change caused by increasing $n_{\mathrm{e}}$
by 2 standard deviations is less pronounced, with regard to both
lower and upper limits on $\lambda_{0}$, as demonstrated in the
right plot in top row of Fig.~\ref{fi:posterior}. 

The middle row of Fig.~\ref{fi:posterior} shows the effect of
including our prior information on $\Omega_{0}$ (see
Sect.~\ref{olprior}).  As is to be expected, (for both values of
$n_{\mathrm{e}}$) lower values of $\Omega_{0}$ are favoured.  This
has the side effect of weakening our lower limit on $\lambda_{0}$
(though only slightly affecting the upper limit). 

We believe that the left plot of the bottom row of
Fig.~\ref{fi:posterior} represents very robust constraints in the
$\lambda_{0}$-$\Omega_{0}$ plane.  The upper limits on
$\lambda_{0}$ come from gravitational lensing statistics, which,
due to the extremely rapid increase in the optical depth for
larger values of $\lambda_{0}$, are quite robust and relatively
insensitive to uncertainties in the input data (compare the left
and right columns of Fig.~\ref{fi:posterior}) as well as to the
prior information used data (compare the upper, lower and middle
rows of Fig.~\ref{fi:posterior}).  The upper and lower limits on
$\Omega_{0}$ are based on a number of different methods and appear
to be quite robust, as discussed in Sect.~\ref{olprior}.  The
combination of the relatively secure knowledge of $H_{0}$ and the
age of the universe combine with lens statistics to produce a good
lower limit on $\lambda_{0}$, although this is to some extent
still subject to the caveats mentioned above. 

\begin{figure*}
   \resizebox{0.375\textwidth}{!}{\includegraphics{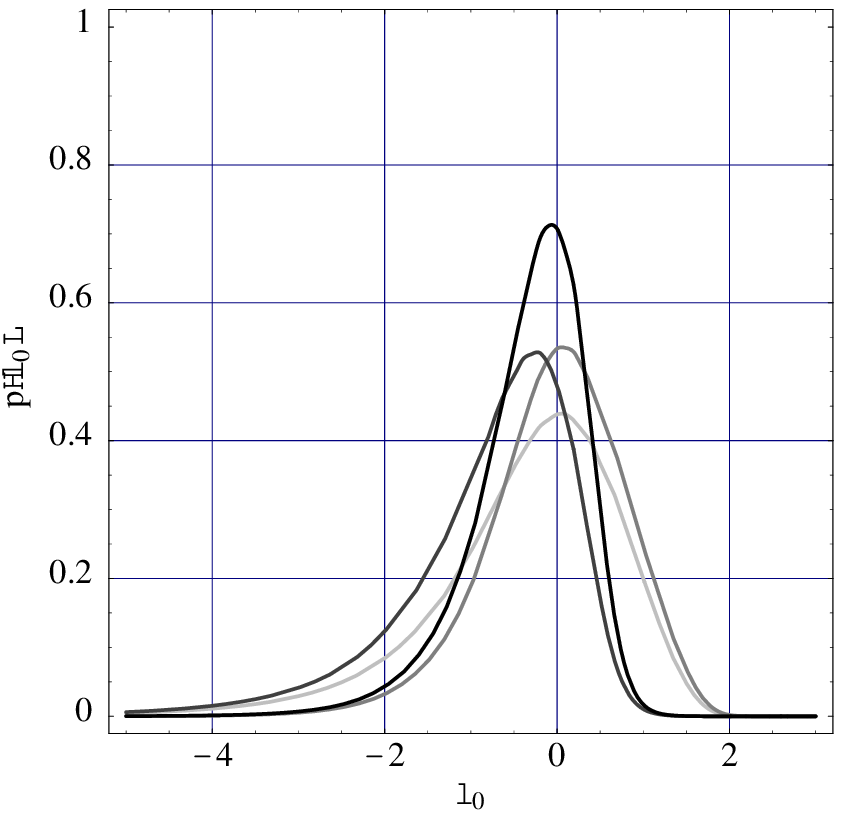}}
   \hfill
   \resizebox{0.375\textwidth}{!}{\includegraphics{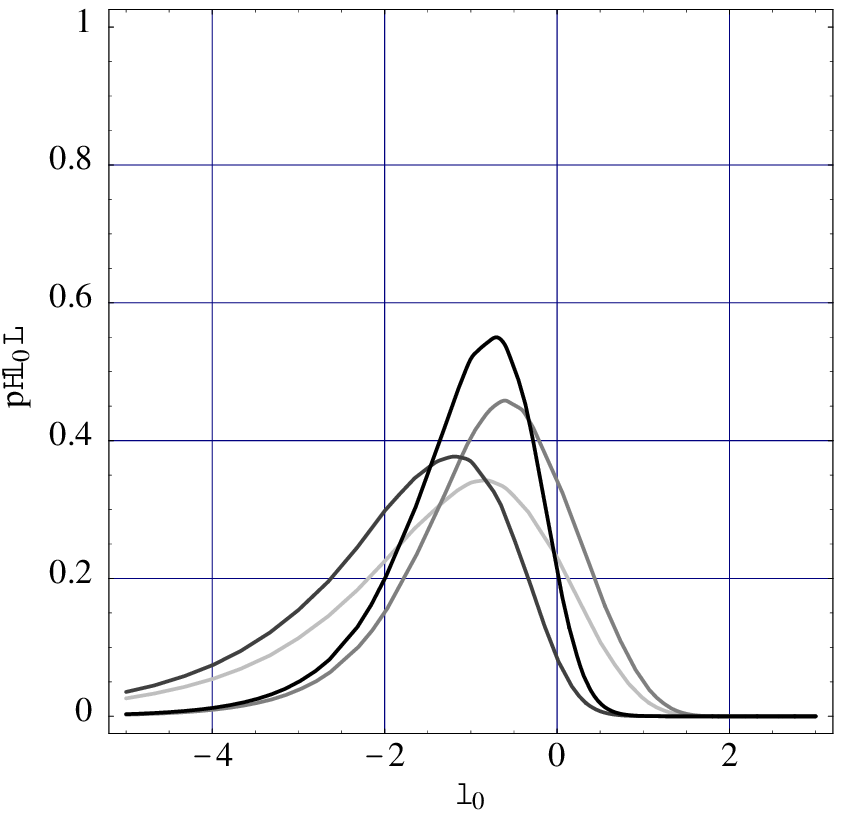}}

   \noindent
   \resizebox{0.375\textwidth}{!}{\includegraphics{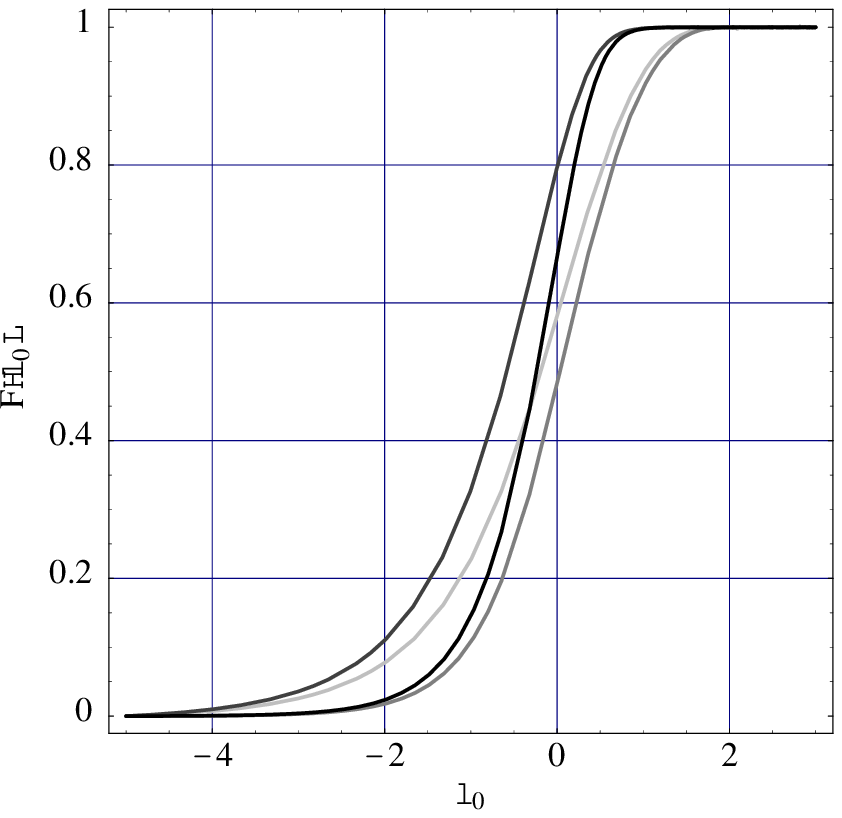}}
   \hfill
   \resizebox{0.375\textwidth}{!}{\includegraphics{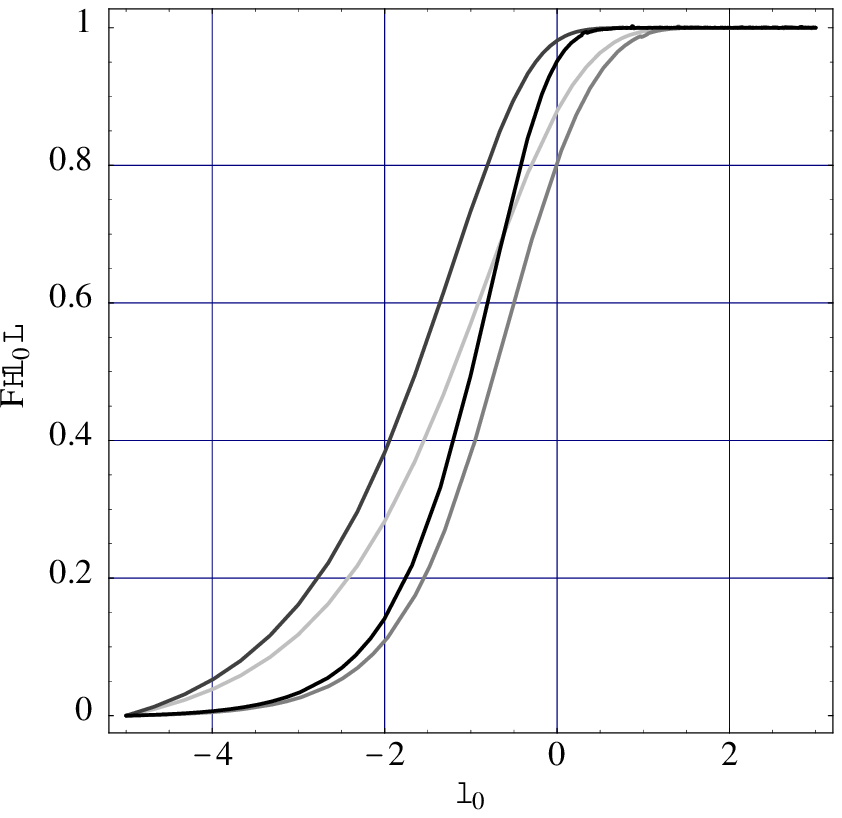}}
   \hfill
   \caption[]{
      \emph{Left column:}
      The top panel shows the normalised marginal likelihood function
      $p(\lambda_0|D)$ (light grey curve) and the marginal posterior
      probability density functions $p_1(D|\lambda_0)$ (medium grey
      curve), $p_2(D|\lambda_0)$ (dark grey curve) and
      $p_3(D|\lambda_0)$ (black curve).  All nuisance parameters are
      assumed to take precisely their mean values.  The bottom panel
      shows the respective cumulative distribution functions.  These can
      be used to construct any desired $\Omega_{0}$-averaged upper or
      lower limits on $\lambda_{0}$. 
      \emph{Right column:} Exactly the same as the left column, but the 
      parameter $n_{\mathrm{e}}$ is increased by two standard deviations} 
   \label{fi:marginal}%
\end{figure*}

If one is interested in the allowed range of $\lambda_{0}$, one
can marginalise over $\Omega_{0}$ to obtain a probability
distribution for $\lambda_{0}$.  This is illustrated in
Fig.~\ref{fi:marginal} 
and Table~\ref{ta:results}.
\begin{table*}
   \caption[]{
      Marginal mean values, standard deviations and $0.95$ confidence
      intervals for the parameter~$\lambda_0$ on the basis of the
      marginal distributions shown in the top row of
      Fig.~\ref{fi:marginal}; `information' refers to Eq.~\ref{eq:info}}
   \label{ta:results}
   \begin{tabular*}{\linewidth}{@{\extracolsep{\fill}}lccccc}
      \hline
      \hline
      Distribution & Mean & standard deviation  &
      \multicolumn{2}{l}{95\% c.l.~range}  & information \\
      \hline
      $p(D|\lambda_0)$   & $-0.35$  & $1.07$ & $-2.55$ & $1.51$  & \\
      $p_1(\lambda_0|D)$ & $-0.02$  & $0.80$ & $-1.59$ & $1.50$ & $1.74$  \\
      $p_2(\lambda_0|D)$ & $-0.78$  & $0.97$ & $-2.85$ & $0.76$ & $1.24$ \\
      $p_3(\lambda_0|D)$ & $-0.34$  & $0.67$ & $-1.72$ & $0.79$ & $1.74$  \\
      \hline
   \end{tabular*}
\end{table*}

\subsection{Comparison with other results}
\label{compare}

For comparison with other results, as a first step one can examine
the allowed range of $\lambda_{0}$ for the current `best-fit'
value for $\Omega_{0}$, which we take, based on the work cited in
Sect.~\ref{olprior}, to be $\Omega_{0}=0.3$.  (A more conservative
estimate is reflected by using the prior probability distribution
$p_2(\lambda_0,\Omega_0) = L(\Omega_0|0.4,0.2)$ as shown by the
dark grey curve in Fig.~\ref{fi:marginal} and in
Table~\ref{ta:results}.) On the other hand, previous limits on
$\lambda_{0}$ have often been quoted for a flat universe (K96 and
references therein).  We consider both cases in
Tables~\ref{ta:specialo} and \ref{ta:specialk}. 
\begin{table*}
\begin{minipage}[t]{\textwidth}
   \caption[]{
      Mean values and ranges for assorted confidence levels for the
      parameter $\lambda_{0}$ for our a priori and various a posteriori
      likelihoods from this analysis and from other tests from the
      literature (using the latest publicly available results) for the
      special case $\Omega_{0} = 0.3$.  Except where noted, the ranges
      quoted are the projections of the corresponding confidence
      contours in the $\lambda_{0}$-$\Omega_{0}$ plane onto the
      $\lambda_{0}$ axis\footnote{Note that some references quote
      confidence ranges for $k = 0$---in general, these will be
      different than the projection of the intersection of the
      corresponding contour in the $\lambda_{0}$-$\Omega_{0}$ plane onto
      the $\lambda_{0}$-axis.} (as opposed to $\Omega_{0}$-independent
      estimates, which of course would always give a smaller range), and
      are of course two-sided, not one-sided, bounds.  Values are either
      those quoted in the references given and/or obtained from figures
      in those references; inequalities mean that the corresponding
      confidence contour is to be found in the range indicated by the
      inequality, e.g.~$<-1.2$ would mean that the corresponding contour
      level is to be found at $\lambda_{0}<-1.2$, \emph{not} that the
      constraint is $\lambda_{0}<-1.2$ at the corresponding confidence
      level.  This arises because the corresponding area of parameter
      space was not examined in the reference in question.  If the
      confidence interval could not be determined from the reference,
      \emph{both} values in the corresponding column are missing} 
   \label{ta:specialo}
   \begin{tabular*}{\linewidth}{@{\extracolsep{\fill}}lrrrrrrrr}
   \hline
   \hline
   Cosmological test& 
   \multicolumn{2}{c}{68\% c.l.~range}  &
   \multicolumn{2}{c}{90\% c.l.~range}  &
   \multicolumn{2}{c}{95\% c.l.~range}  &
   \multicolumn{2}{c}{99\% c.l.~range}  \\
   \hline
this work, $p(D|\lambda_0)$ &
$-1.18$ & $0.24$ & $-2.19$ & $0.50$ & $-2.81$ & $0.60$ & $-4.16$ & $0.73$ \\
this work, $p_1(\lambda_0|D)$ & 
$-0.97$ & $0.46$ & $-1.55$ & $0.60$ & $-1.89$ & $0.69$ & $-2.73$ & $0.81$ \\ 
this work, $p_2(\lambda_0|D)$ &
$-2.00$ & $0.49$ & $-3.33$ & $0.65$ & $-4.10$ & $0.72$ & $<-5.00$ & $0.80$ \\ 
this work, $p_3(\lambda_0|D)$ & 
$-1.20$ & $0.52$ & $-1.98$ & $0.69$ & $-2.35$ & $0.77$ & $-3.40$ & $0.86$ \\ 
lens statistics (K96) & 
\multicolumn{8}{c}{not possible since only $ k=0 $ models considered} \\
radio lenses\footnote{\citet{EFalcoKM98a}}\footnote{contour at 95.4\% not 95\%} & 
$-0.54$ & $0.28$ & $<-1.00$ & $0.75$ & $<-1.00$ & $0.89$& $$ & $$ \\ 
optical lenses\footnote{\citet{EFalcoKM98a}}\footnote{contour at 95.4\% not 95\%} & 
$<-1.00$ & $0.37$ & $<-1.00$ & $0.75$ & $<-1.00$ & $0.89$ & $$ & $$ \\ 
radio + optical lenses\footnote{\citet{EFalcoKM98a}}\footnote{contour at 95.4\% not 95\%} & 
$<-1.00$ & $-0.12$ & $<-1.00$ & $0.50$ & $<-1.00$ & $0.70$ & $<-1.00$ & $0.89$ \\ 
supernovae $m$-$z$ relation $\cal A$\footnote{\citet{SPerlmutteretal98a}} &
$-0.70$ & $0.50$ & $-1.15$ & $0.75$ & $$ & $$ & $$ & $$ \\ 
supernovae $m$-$z$ relation $\cal B$\footnote{\citet{ARiessetal98a}}\footnote{Fig.~6, solid contours}\footnote{contours at 68.3\%, 95.4\% and 99.7\% instead of 68\%, 95\% and 99\% respectively} &
$0.78$ & $1.00$ & $$ & $$ & $0.53$ & $1.27$ & $0.27$ & $1.41$ \\ 
CNOC survey\footnote{\citet{RCarlberg98a}} &
$<-0.50$ & $<-0.50$ & $<-0.50$ & $<-0.50$ & $$ & $$ & $$ & $$ \\ 
CMB\footnote{\citet{CLineweaver98a}}\footnote{contours at 68.3\%, 95.4\% and 99.7\% instead of 68\%, 95\% and 99\%, respectively}  &
$0.44$ & $0.67$ & $$ & $$ & $0.36$ & $>0.90$ & $0.26$ & $>0.90$ \\ 
CMB + \textit{IRAS}\footnote{\citet{MWebsterBHLLR98a}} &
\multicolumn{8}{c}{not possible since only $ k=0 $ models considered} \\
double radio sources\footnote{\citet{EGuerraDW98a}} &
$0.00$ & $1.00$ & $<-2.00$ & $1.39$ & $$ & $$ & $$ & $$ \\ 
   \hline
   \end{tabular*}
   \end{minipage}
\end{table*}
\begin{table*}
\begin{minipage}[t]{\textwidth}
   \caption[]{
      Mean values and ranges for assorted confidence levels for the
      parameter $\lambda_{0}$ for our a priori and various a posteriori
      likelihoods from this analysis and from other tests from the
      literature (using the latest publicly available results) for the
      special case $k = 0$.  Otherwise the same as
      Table~\ref{ta:specialo}, in particular the references are not
      listed in the footnotes to this table.  $X$ denotes the fact that
      there is no intersection of the confidence contour with the
      $k=0$ line} 
   \label{ta:specialk}
   \begin{tabular*}{\linewidth}{@{\extracolsep{\fill}}lrrrrrrrr}
   \hline
   \hline
   Cosmological test& 
   \multicolumn{2}{c}{68\% c.l.~range}  &
   \multicolumn{2}{c}{90\% c.l.~range}  &
   \multicolumn{2}{c}{95\% c.l.~range}  &
   \multicolumn{2}{c}{99\% c.l.~range}  \\
   \hline
this work, $p(D|\lambda_0)$ &
$-0.68$ & $0.51$ & $<-1.00$ & $0.57$ & $<-1.00$ & $0.62$ & $<-1.00$ & $0.70$ \\ 
this work, $p_1(\lambda_0|D)$ & 
$-0.09$ & $0.56$ & $-0.38$ & $0.64$ & $-0.57$ & $0.68$ & $-1.04$ & $0.81$ \\ 
this work, $p_2(\lambda_0|D)$ &
$X$ & $X$ & $0.09$ & $0.69$ & $-0.03$ & $0.73$ & $-0.28$ & $0.92$ \\ 
this work, $p_3(\lambda_0|D)$ & 
$0.47$ & $0.48$ & $0.18$ & $0.67$ & $0.07$ & $0.70$ & $-0.14$ & $0.84$ \\ 
lens statistics\footnote{value for $k=0$, not projection} &
$$ & $$ & $$ & $$ & $<0.00$ & $0.66$ & $$ & $$ \\ 
radio lenses\footnote{contour at 95.4\% not 95\%} & 
$-0.47$ & $0.56$ & $<-1.00$ & $0.72$ & $<-1.00$ & $0.80$ & $<-1.00$ & $0.85$ \\ 
optical lenses\footnote{contour at 95.4\% not 95\%} & 
$<-1.00$ & $0.56$ & $<-1.00$ & $0.72$ & $<-1.00$ & $0.80$ & $<-1.00$ & $0.87$ \\ 
radio + optical lenses\footnote{contour at 95.4\% not 95\%} & 
$-0.87$ & $0.43$ & $<-1.00$ & $0.60$ & $<-1.00$ & $0.69$ & $<-1.00$ & $0.78$ \\ 
supernovae $m$-$z$ relation $\cal A$ &
$0.20$ & $0.60$ & $-0.05$ & $0.75$ & $$ & $$ & $$ & $$ \\ 
supernovae $m$-$z$ relation $\cal B$\footnote{Fig.~6, solid contours}\footnote{contours at 68.3\%, 95.4\% and 99.7\% instead of 68\%, 95\% and 99\% respectively} &
$0.74$& $0.83$ & $$ & $$ & $0.61$ & $0.92$ & $0.50$ & $1.00$ \\ 
CNOC survey &
$0.85$ & $0.95$ & $0.81$ & $0.98$ & $$ & $$ & $$ & $$ \\ 
CMB\footnote{contour at 68.3\% instead of 68\%; other contours, and part
of the 68.3\% contour, lie partially in 
the $k=+1$ area of parameter space which was not examined for technical 
reasons in \citet{CLineweaver98a}}  &
$<0.00$ & $0.60$ & $<0.00$ & $<0.00$ & $<0.00$ & $<0.00$ & $<0.00$ & $<0.00$ \\ 
CMB + \textit{IRAS}\footnote{value for $k=0$, not projection} &
$0.47$ & $0.71$ & $$ & $$ & $$ & $$ & $$ & $$ \\ 
double radio sources &
$0.35$ & $1.00$ & $0.70$ & $1.00$ & $$ & $$ & $$ & $$ \\ 
\hline
\end{tabular*}
\end{minipage}
\end{table*}

We do not do a comparison for the special case $\lambda_{0} = 0$
since this analysis of gravitational lensing statistics does not
usefully constrain $\Omega_{0}$ (any limits coming only from the
prior information on $\Omega_{0}$). 

It is beyond the scope of this paper to do a full comparison of
different cosmological tests.  Except for a few general comments,
we therefore restrict ourselves to comments on the similarities
and differences between the results from this work without using
prior information on $\lambda_{0}$ and $\Omega_{0}$, i.e.~(the
left plot in) Fig.~\ref{fi:likelihood}, and the those from K96 and
\citet{EFalcoKM98a} (using only optical data, i.e.~the lower left
plot in their Fig.~5). 

Taking all results at face value and examining the $\Omega_{0} = 0.3$
case first, we note that with `three-and-one-half' exceptions (counting
as one test each the four from this work and the three from
\citet{EFalcoKM98a}) the 68\% c.l.~\emph{lower} limit from
\citet{CLineweaver98a} is \emph{higher} than \emph{all} 68\%
\emph{upper} limits from other tests, while the 95\% lower and upper
confidence levels from \citet{CLineweaver98a} are higher than the
corresponding limits from the other tests for all but one of these. 
Even at the 99.9\% confidence level (not shown in
Table~\ref{ta:specialo}), the \citet{CLineweaver98a} result requires
$\lambda_{0} \ge 0.12$.  If one assumes $\Omega_{0} = 0.3$, only
\citet{CLineweaver98a} requires $\lambda_{0} >0$, though all other tests
(except \citet{RCarlberg98a}) are compatible with this.  This is not
surprising, since it is well-known that constraints from CMB
anisotropies tend to run more or less orthogonal in the
$\lambda_{0}$-$\Omega_{0}$ plane to those from most other tests
\citep[e.g.][]{MWhite98a,DEisensteinHT98b,MTegmarkEH98a,MTegmarkEHK98a}.

Examining the $k=0$ case, it is interesting to note that the 68\% (90\%)
confidence level \emph{lower} limit on $\lambda_{0}$ from
\citet{RCarlberg98a} is \emph{higher} than \emph{all} of the 68\% (90\%)
c.l.~\emph{upper} limits from \emph{all} other tests except
\citet{EGuerraDW98a}.  Otherwise, with `one-and-one-half' exceptions all
tests are compatible even at the 68\% confidence level.  If one assumes
$k = 0$, then the evidence for $\lambda_{0} > 0$ looks convincing: at
the 68\% confidence level, again with `one-and-one-half' exceptions, all
tests indicate $\lambda_{0} > 0$; even at 90\% the evidence is still
quite good, if one keeps in mind that the gradient towards smaller
values of $\lambda_{0}$ is generally not as steep as towards larger
values. 

Again taken at face value, neither the $k = 0$ case nor the
$\Omega_{0} = 0.3$ case are compatible with all tests, even at the
$\approx$90\% confidence level.  It appears the simplest solution
to achieve concordance would be to have $\Omega_{0} \approx 0.2$,
which is within the error on $\Omega_{0}$ discussed in
Sect.~\ref{olprior}.  For $k=0$ this would imply $\lambda_{0} =
0.8$, which seems to be ruled out, thus ruling out the flat
universe altogether.  For a non-flat universe, reducing
$\Omega_{0}$ would, due to the CMB constraint, require a higher
value of $\lambda_{0}$, and thus make the $\lambda_{0} = 0$ case
more unlikely, ruling out this special case as well. 

On balance, a cosmological model with $\lambda_{0} \approx 0.3$
and $\Omega_{0} \approx 0.25$ seems compatible with all known
observational data (not just those discussed here) at a
comfortable confidence level. 

\begin{figure*}
   \resizebox{0.375\textwidth}{!}{\includegraphics{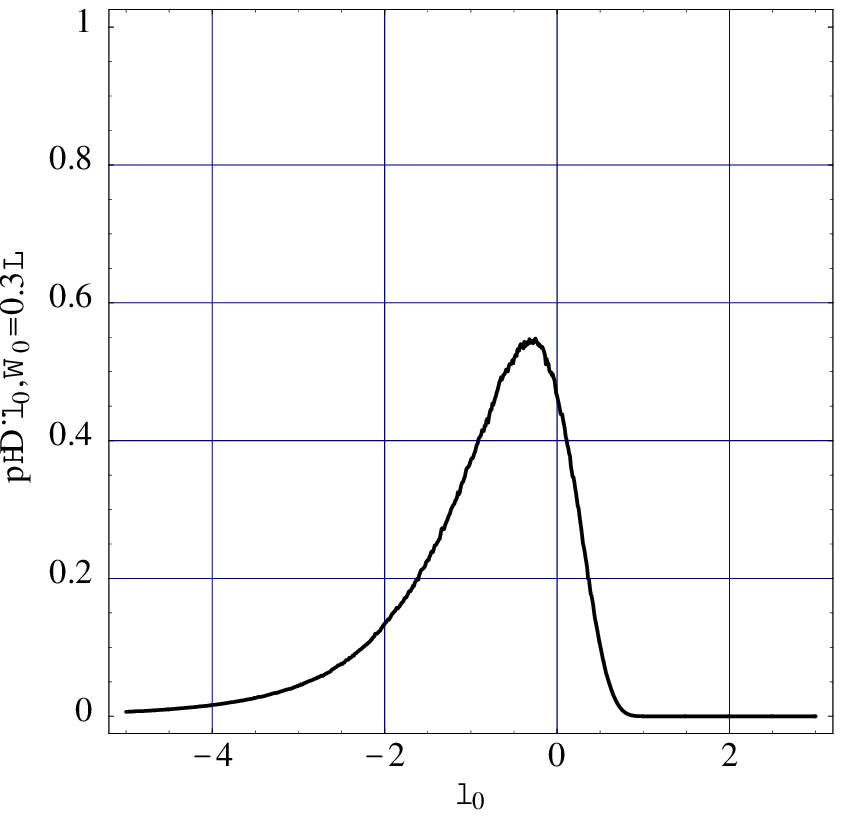}}
   \hfill
   \resizebox{0.375\textwidth}{!}{\includegraphics{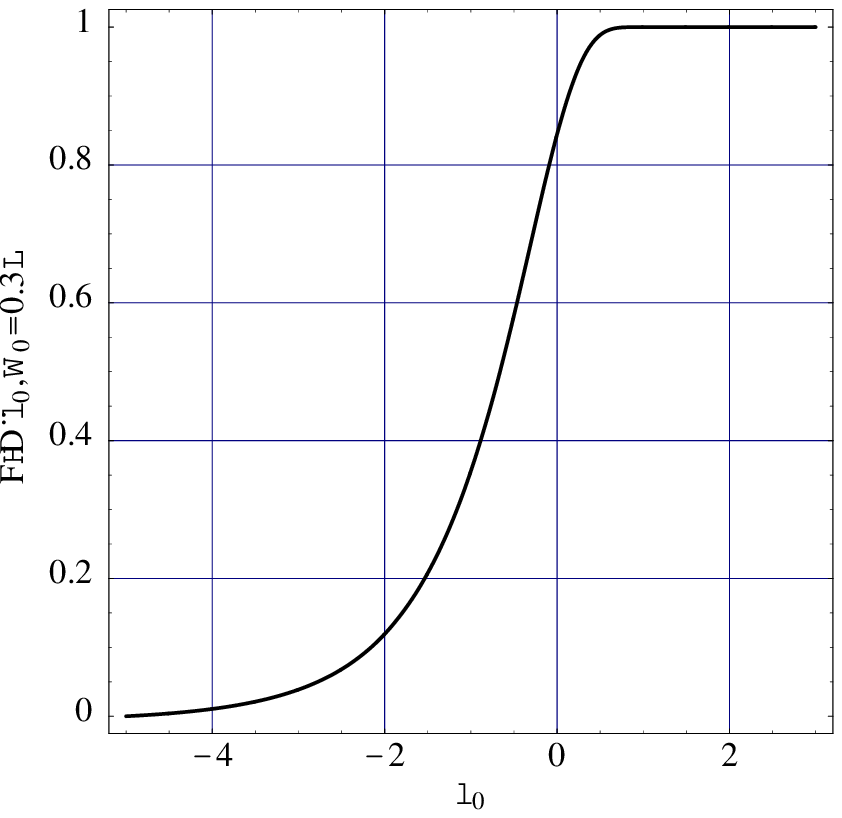}}
   \caption[]{
      \emph{Left panel:} The likelihood function as a function of
      $\lambda_{0}$ for $\Omega_{0}=0.3$ and with all nuisance
      parameters taking their default values.  \emph{Right panel:} The
      same but plotted cumulatively.  See Table~\ref{ta:03}} 
   \label{fi:03}
\end{figure*}

For a `likely' $\Omega_{0}$ value of 0.3 we have calculated the
likelihood with the higher resolution $\Delta\lambda_{0}=0.01$. 
This is shown in Fig.~\ref{fi:03}. 
From these calculations one can extract confidence limits which,
due to the higher resolution in $\lambda_{0}$, are more accurate. 
These are presented in Table~\ref{ta:03} and should be compared to
those for $p(D|\lambda_0)$ from Table~\ref{ta:specialo}. 
\begin{table*}
   \caption[]{
      Confidence ranges for $\lambda_{0}$ assuming $\Omega_{0}=0.3$.
      Unlike the results presented in Table~\ref{ta:specialo}, these
      figures are for a specific value of $\Omega_{0}$ and not the
      values of intersection of particular contours with the
      $\Omega_{0}=0.3$ line in the $\lambda_{0}$-$\Omega_{0}$ plane.
      These are more appropriate if one is convinced that
      $\Omega_{0}=0.3$ and have been calculated using ten times better
      resolution than the rest of our results presented in this work. 
      See Fig.~\ref{fi:03}}    
   \label{ta:03} 
   \begin{tabular*}{\linewidth}{@{\extracolsep{\fill}}rrrrrrrr}
   \hline
   \hline
   \multicolumn{2}{c}{68\% c.l.~range}  &
   \multicolumn{2}{c}{90\% c.l.~range}  &
   \multicolumn{2}{c}{95\% c.l.~range}  &
   \multicolumn{2}{c}{99\% c.l.~range}  \\
   \hline
$-1.27$ & $0.27$ & $-2.26$ & $0.51$ & $-2.87$ & $0.60$ & $-4.10$ & $0.72$ \\ 
   \hline
   \end{tabular*}
\end{table*}

Again, a full discussion of joint constraints involving discussion
of possible sources of error for each test, as well as comparing
the full contours in the $\lambda_{0}$-$\Omega_{0}$ plane, is
beyond the scope of this paper.  However, quick comparisons would
be aided were the results of all tests available in an
easy-to-process electronic form (see below); such quick
consistency tests would enable one to spot areas of inconsistency
much more quickly.  Also, it should be emphasised that the
projections onto the $\lambda_{0}$-axis of the intersection of a
particular confidence contour with the $\Omega_{0}=0.3$ or $k=0$
axis are generally \emph{not} the same as the corresponding
confidence interval for the $\Omega_{0}=0.3$ or $k=0$ special
cases. 

For a flat universe, our 95\% confidence level upper limit on
$\lambda_{0}$-$\Omega_{0}$, i.e.~the value of $\lambda_{0}$ where
this contour crosses the $k = 0$ line, is $\lambda_{0} < 0.62$. 
This is essentially the same as the $\lambda_{0} < 0.66$ of K96,
as was to be expected considering we used essentially the same
data and methods.  Interpreted cautiously, one might conclude from
this that the singular isothermal sphere model is a good
approximation as far as determining the cosmological parameters
from lens statistics is concerned, as was assumed in
\citet{EFalcoKM98a}. Our 99\% confidence level upper limit on
$\lambda_{0}$ is $\lambda_{0} < 0.70$.  This is quite a tight
upper bound on $\lambda_{0}$ and appears to be quite robust. 

Perhaps more interesting is the comparison with (the results using
only optical data in) \citet{EFalcoKM98a}.  Although a detailed
comparison is complicated by the different plotting scheme and
reducing the entire contour (or indeed grey-scale) plot to a few
numbers throws away information, it is obvious that the plots are
broadly similar.  Our 68\% contour is, for $\Omega_{0} \approx 1$,
roughly parallel to the $\lambda_{0}$-axis at $\lambda_{0}\approx
-1$.  This is just at the edge of the \citet{EFalcoKM98a} plot,
and as they provide no grey-scale, it is difficult to compare the
lower limits on $\lambda_{0}$.  Thus, while our main goal was to
explore a `large enough' region of parameter space, comparison in
the areas where there is overlap shows consistency, which
strengthens our faith in the conclusions pertaining to areas of
parameter space where there is no overlap. 

Recently, it has become quite fashionable to discuss joint
constraints derived from a variety of cosmological tests.  This
has grown from plotting the overlap of likelihood contours (often
in a space spanned by parameters other than $\lambda_{0}$ and
$\Omega_{0}$)
\citep[e.g.][]{JOstrikerPSteinhardt95a,MTurner96a,JBaglaPN96a,LKrauss98a,
MWhite98a} to full-blown joint likelihood analyses, both detailed
theoretical investigations of what will be possible in the future
\citep[e.g.][]{MTegmarkEH98a,MTegmarkEHK98a,DEisensteinHT98a,
DEisensteinHT98b} and more restricted analyses using present data
\citep[e.g.][]{MWebsterBHLLR98a}.  While in some cases it is quick
and easy to calculate the likelihood as a function of
$\lambda_{0}$ and $\Omega_{0}$ given the data, for example for
tests using the $m$-$z$ relation, in other cases such as the
present one it is a major programming and computational effort to
do so.  To aid comparisons, all figures from this paper are
available in the form of tables of numbers at 
\begin{quote}
\verb|http://multivac.jb.man.ac.uk:8000/ceres|\\
                                       \verb|/data_from_papers/lower_limit.html|
\end{quote}
and we urge our colleagues to follow our example.  We applaud the
fact that most results are now presented in the
$\lambda_{0}$-$\Omega_{0}$ plane, as opposed to using other
parameters such as $q_{0}$ or
$\Omega_{\mathrm{tot}}\equiv\lambda_{0}+\Omega_{0}$.  A further
aid in comparison would be a uniform choice of axes.  We prefer to
plot $\Omega_{0}$ on the $y$-axis and $\lambda_{0}$ on the $x$
axis since up/down symmetry is less fundamental than left/right
symmetry and this mirrors the fact that $\Omega_{0}$ has the
physical lower limit $\Omega_{0} = 0$ whereas no corresponding
upper or lower limits for $\lambda_{0}$ exist.  Square plots with
the same range would further aid the comparison.  Of course, if
all data are publicly available, then they can be re-plotted to
taste.

\section{Summary and conclusions}
\label{conclusions}

We have re-analysed optical gravitational lens surveys from the
literature, using the techniques described in
\citet{CKochanek96a}, for the first time allowing both the
cosmological constant $\lambda_{0}$ and the density parameter
$\Omega_{0}$ to be free parameters while also using a non-singular
lens model.  We confirm the well-known results that gravitational
lensing statistics can provide a good upper limit on $\lambda_{0}$
but are relatively insensitive to $\Omega_{0}$.  We have presented
the new result of a robust lower limit on $\lambda_{0}$, which is
a substantial improvement on previously known \emph{robust} lower
limits.  Coupled with relatively conservative prior information
about the Hubble constant $H_{0}$, the age of the universe and the
well-established value of $\Omega_{0}$, one can reduce the allowed
parameter space in the $\lambda_{0}$-$\Omega_{0}$ plane to a
small, finite region, which is similar to the area allowed by
joint constraints based on many other cosmological tests (see
Fig.~\ref{fi:posterior}).

Using lens statistics information alone, at 95\% confidence, our
lower and upper limits on $\lambda_{0}-\Omega_{0}$ are
respectively $-3.17$ and $0.3$.  For a flat universe, this
corresponds to lower and upper limits on $\lambda_{0}$ of
respectively $-1.09$ and $0.65$.  Keeping in mind the difficulties
of a quantitative comparison, this is in good agreement with other
recent measurements of the cosmological constant. This value was
calculated from Table~\ref{ta:03} and assuming a degeneracy in
$\lambda_{0} - \Omega_{0}$ as in \citet{ACoorayQC99a} and
\citet{ACooray99a}. For comparison, from Table~\ref{ta:specialk},
the corresponding value for the upper limit on $\lambda_{0}$ is
$0.62$ and the value from K96 is $0.66$.\footnote{The value from
\citet{ACoorayQC99a}  and \citet{ACooray99a} is 0.79, but it
should be noted this value (the same in both papers) is based on
different surveys, namely the Hubble Deep Field and CLASS,
respecively.)} 

For detailed comparison of cosmological tests, one needs to
compare confidence contours---calculated in the same, preferably
in the `real', way---in the same parameter space.  Of course, this
makes it difficult to meaningfully reduce the results of a given
cosmological test to one or even a few single numbers.  Unless a
cosmological test is developed which can measure $\lambda_{0}$
independently of any other parameters, there is not much point in
quoting unqualified `limits on $\lambda_{0}$'. 

Presently tentative claims of the detection of a positive
cosmological constant, if true, would rank among the great
discoveries of cosmology. Even though there are serious
difficulties involved, it seems worthwhile to be able to confirm
this result by improving the lower limit on $\lambda_{0}$ derived
from gravitational lensing statistics.  Targetting the two primary
sources of uncertainty calls for improving our knowledge of the
normalisation of the local luminosity density of galaxies as well
as increasing the size of gravitational lens surveys.  As far as
the latter goes, the CLASS survey
\citep{IBrowneJAHMNWBKBFBPRWP97a,SMyersetal99a} looks the most
promising at the moment.  In a companion paper
\citep{PHelbigMQWBK99a}, we have shown that comparable
constraints to the ones presented in this work can be obtained
from the JVAS gravitational lens survey; this gives us hope that
the much larger CLASS survey will offer improvement in this area. 

Cosmological tests which set tight upper limits on $\Omega_{0}$
imply, for a flat $k = 0$ universe, a value of $\lambda_{0}$ which
is ruled out by lensing statistics.  For a non-flat universe, many
tests are indicating $\lambda_{0} > 0$, and at present a
cosmological model with $\lambda_{0} \approx 0.3$ and $\Omega_{0}
\approx 0.25$ seems compatible with all known observational data,
with neither a flat universe nor a universe without a positive
cosmological constant being viable alternatives.  The simplest
case, the Einstein-de~Sitter universe with $\lambda_{0} = 0$ and
$\Omega_{0} = 1$, both flat and without a cosmological constant,
had been abandoned long before the new observational data cited in
this work came to light \citep[see, e.g.,][and references
therein]{JOstrikerPSteinhardt95a}; this trend has continued, with
the next-most-simple cases also no longer viable.  For
$\lambda_{0}$ and $\Omega_{0}$, we have in a sense reached the
least simple case; it will be interesting to see if this trend
continues with regard to the other cosmological parameters, in
particular those which can be measured by the \textit{Planck
Surveyor} mission.  Larger gravitational lens surveys such as CLASS
will be a step in this direction.

\begin{acknowledgements}
We thank Sjur Refsdal, Leon Koopmans, Lutz Wisotzki and many
colleagues at Jodrell Bank for helpful comments and suggestions.
This research was supported in part by the European Commission,
TMR Programme, Research Network Contract ERBFMRXCT96-0034 `CERES'.
\end{acknowledgements}

\appendix

\section{Getting a feel for it}

\begin{figure}
   \resizebox{0.375\textwidth}{!}{\includegraphics{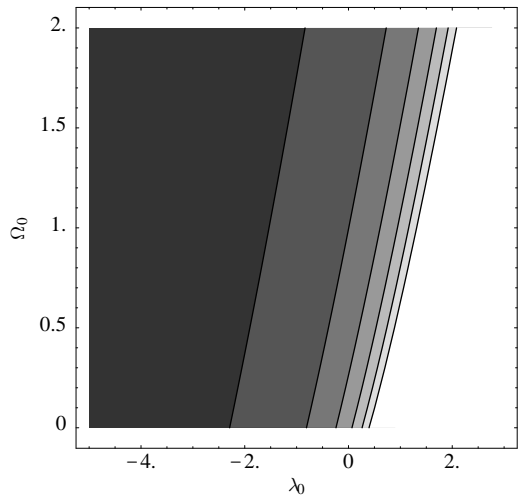}}
   \caption[]{
      Likelihood that the non-lenses in our sample are not lenses.
      The contour levels mark changes of a factor of ten in the probability, 
      which is also indicated by the grey scale, darker values corresponding 
      to higher values}
   \label{fi:nonlenses}
\end{figure}

\begin{figure}
   \resizebox{0.375\textwidth}{!}{\includegraphics{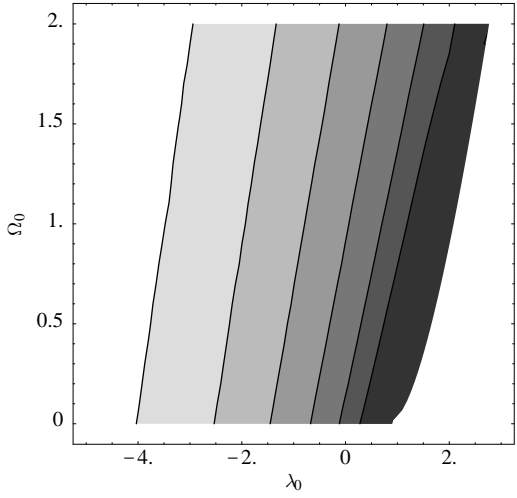}}
   \caption[]{
      Likelihood that the lenses in our sample have the properties they
      do.  The contour levels mark changes of a factor of ten in the
      probability, which is also indicated by the grey scale, darker
      values corresponding to higher values} 
   \label{fi:lenses}
\end{figure}

\begin{figure}
   \resizebox{0.375\textwidth}{!}{\includegraphics{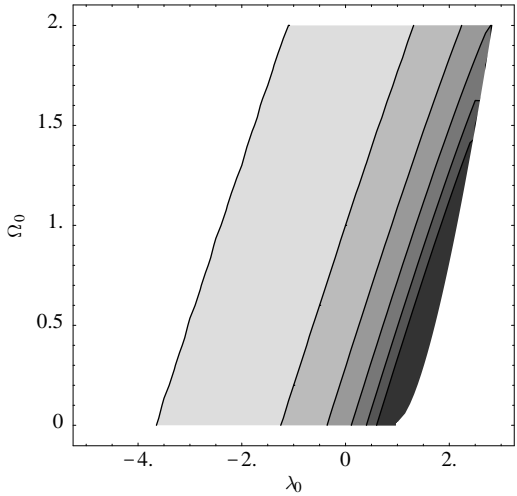}}
   \caption[]{
      The volume element $\mathrm{d}V/\mathrm{d}z$ at the typical lens
      galaxy redshift $z_{\mathrm{d}} = 0.7$.  The contours indicate the
      fraction $0.1,0.2,\dots0.6$ of the volume element in the limiting
      case of the de~Sitter model ($\lambda_{0} = 1$, $\Omega_{0} = 0$).
      This is also indicated by the grey scale, darker values
      corresponding to a larger volume.  For smaller redshifts the
      contours are more vertical (and further apart), for larger
      redshifts more horizontal (cf.~Fig.~3 of \citet{MTegmarkEHK98a}
      but note their swapped axes)} 
   \label{fi:volume}
\end{figure}

\begin{figure}
   \resizebox{0.375\textwidth}{!}{\includegraphics{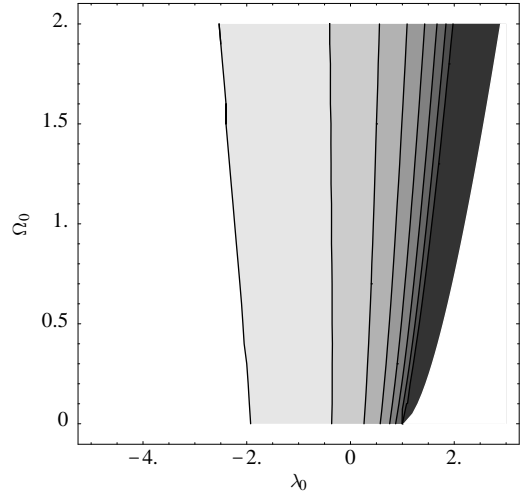}}
   \caption[]{
      Cross section for the softened singular isothermal sphere
      model used in this work for a typical lens redshift
      $z_{\mathrm{d}} = 0.7$ and a typical source redshift
      $z_{\mathrm{s}} = 2.0$ for the fiducial values $\sigma=\sigma_{*}$
      and $s=s{*}$ (see Sect.~\ref{calculations}).  The contours indicate
      the fraction $0.3,0.4,\dots1.0$ of the cross section in the
      limiting case of the de~Sitter model ($\lambda_{0} = 1$,
      $\Omega_{0} = 0$).  This is also indicated by the grey scale,
      darker values corresponding to a larger cross section} 
   \label{fi:crosssection}
\end{figure}

\begin{figure}
   \resizebox{0.375\textwidth}{!}{\includegraphics{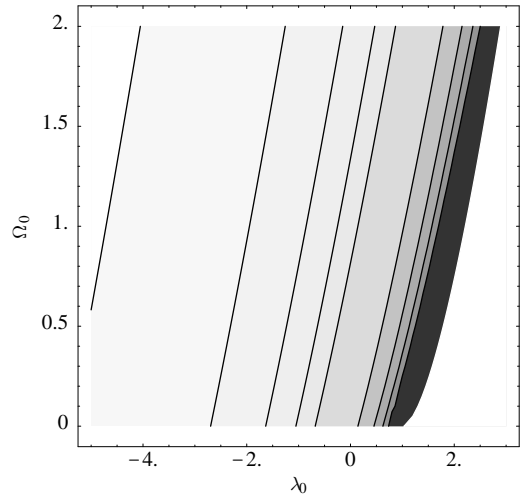}}
   \caption[]{
      Expected number of lenses.  Contours, from left to right,
      indicate 1, 2, 3, 4, 5, 10, 15, 20 and 25 lenses.  Darker values
      of the grey scale correspond to higher values.
      Cf.~\citet{ACoorayQC99a} and \citet{ACooray99a} where the number
      of lenses as a function of $\lambda_{0}$ and $\Omega_{0}$ has been
      calculated for the Hubble Deep Field and for CLASS} 
   \label{fi:number}
\end{figure}

The likelihood of a given cosmological model for a given set of
observational data, calculated using Eq.~(\ref{eq:p}), is the
result of the complex interplay of many factors.  While this is
necessary for a detailed analysis, it perhaps obscures the fact
that the likelihood is basically the product of two terms, the
likelihood that the non-lenses in our sample are not lenses (see
Fig.~\ref{fi:nonlenses}) 
and the likelihood that the lenses in our sample (see Fig.~\ref{fi:lenses})
have the observed properties.\footnote{It is interesting to note
that the measurement of $\lambda_{0}$ by \citet{MImGR97a} (who
obtain $\lambda_{0} = 0.64^{+0.15}_{-0.26}$ for a flat universe
and thus a lower limit) essentially corresponds to
Fig.~\ref{fi:lenses} (though with a different sample of lenses). 
Since lensing is a rare phenomenon, small-number statistics are a
source of concern.  The advantage of a well-defined
\emph{gravitational lens survey}, as opposed to using a `sample
from the literature', is that the (much greater) number of
non-lenses in the sample also makes a contribution.  A comparison
of Figs.~\ref{fi:nonlenses} and \ref{fi:lenses} hints that not
taking the non-lenses into account would tend to favour a high
value of $\lambda_{0}$, as indeed found by \citet{MImGR97a}.} The
latter in turn is the result of two basic effects: the dependency
of the volume element $\mathrm{d}V/\mathrm{d}z$ on $\lambda_{0}$
and $\Omega_{0}$ (see Fig.~\ref{fi:volume}) 
and the dependency on the lensing cross section on $\lambda_{0}$
and $\Omega_{0}$ (see Fig.~\ref{fi:crosssection}). 
One can also use the probability that the non-lenses in our sample
are not lenses (illustrated in Fig.~\ref{fi:nonlenses}) to
calculate the expected number of lenses in our sample (see
Fig.~\ref{fi:number}), 
although obviously just counting the number of lenses does not
make use of as much of the available information as does using
Eq.~(\ref{eq:p}).

\bibliographystyle{aa}

\end{document}